\documentclass{aa}
\usepackage[varg]{txfonts}

\usepackage{natbib}
\bibpunct{(}{)}{;}{a}{}{,}

\usepackage{threeparttablex}
\usepackage{longtable}
\usepackage{lscape}

\usepackage{verbatim}
\usepackage{makecell}
\usepackage{caption}
\usepackage{subcaption}
\usepackage{multirow}
\usepackage{booktabs}
\usepackage{hyperref}
\usepackage{listings}
\usepackage{xcolor}

\lstdefinelanguage{json}{
    basicstyle=\ttfamily\small,        
    commentstyle=\color{lightgray}, 
    stringstyle=\color{blue},
    showstringspaces=false,
    breaklines=true, 
    morecomment=[l]{\#},
    morestring=[b]",
    escapeinside={(*@}{@*)},
}

\def\teff{$T_{\rm eff}$}
\def\logg{$\log\,g$}

\def\vsini{$v\sin\,i$}

\begin{document}

\title{Observational mapping of the mass discrepancy in eclipsing binaries\thanks{
Based on observations collected 
with the { \sc hermes} spectrograph mounted on the KU Leuven Mercator telescope, La Palma, Spain.}. A new self-contained framework for concurrent analysis of photometric and spectroscopic time series}
\author{Nadya Serebriakova\inst{\ref{KUL}}
\and Andrew Tkachenko\inst{\ref{KUL}}
\and Cole Johnston \inst{\ref{Surrey},\ref{MPA},\ref{KUL}}
\and Kre\v{s}imir Pavlovski\inst{\ref{Zagreb},\ref{KUL}}
\and Conny Aerts\inst{\ref{KUL},\ref{MPIA},\ref{Radboud}}
}

\institute{Institute of Astronomy, KU Leuven, Celestijnenlaan 200D, B-3001 Leuven, Belgium \\ \email{nadya.serebriakova@kuleuven.be} \label{KUL} 
\and Astrophysics group, Department of Physics, University of Surrey, Guildford, GU2 7XH, United Kingdom \label{Surrey}
\and Max Planck Institute for Astrophysics, Karl-Schwarzschild-Straße 1, 85741 Garching, Germany \label{MPA}
\and Department of Physics, Faculty of Science, University of Zagreb, 10\,000 Zagreb, Croatia \label{Zagreb}
\and Max Planck Institute for Astronomy, K\"onigstuhl 17, 69117 Heidelberg, Germany \label{MPIA}
\and Department of Astrophysics, IMAPP, Radboud University Nijmegen, PO Box 9010, 6500 GL Nijmegen, The Netherlands\label{Radboud}
}

\date{Received XXX / Accepted XXX}

\abstract{
The mass discrepancy problem, observed in high-mass stars within eclipsing binaries, highlights systematic differences between dynamical and evolutionary mass estimates, challenging the accuracy of stellar evolution models. 
}{
We aim to determine whether analysis methods directly contribute to this discrepancy and to assess how methodological improvements might reduce or clarify it. 
}{
To address this, we developed a new self-contained framework that simultaneously models the photometric and spectroscopic data, minimising biases introduced by traditional iterative approaches and enabling consistent parameter optimisation. 
}{
We present this framework alongside validation tests on synthetic data and demonstrate its application to three well-studied observed binaries, including one system known for its pronounced mass discrepancy. The framework recovers multiple viable solutions from distinct local minima, including one that reduces the mass discrepancy. 
}{
 These results illustrate how methodological biases, rather than evolutionary model assumptions, can contribute to the mass discrepancy problem. We further highlight that external constraints, such as independent distance estimates or evolutionary models, may be necessary to distinguish between degenerate solutions. Expanding this analysis to a larger sample will provide a more complete understanding, with forthcoming results in the next paper in this series. 
}


\titlerunning{re:pair}
\authorrunning{Serebriakova et al.}
\maketitle

\section{Introduction}\label{sec:introduction}

The mass discrepancy problem in stellar astrophysics has become a persistent challenge in the study of intermediate- to high-mass stars. The issue manifests as a mismatch between dynamical masses derived from orbital solutions for binaries and those predicted by stellar evolution models, also known as evolutionary masses \citep{claret_dependence_2016,claret_dependence_2017,claret_dependence_2018,claret_dependence_2019}. Although initially found in single O-type stars as a mismatch of spectroscopically derived masses and evolutionary ones \citep{herrero_fundamental_2000,herrero_spectroscopic_2007, markova_spectroscopic_2018}, the most concerning are discrepancies of dynamical masses in double-lined eclipsing systems, where both masses and radii can be determined independently from models with a high level of precision and accuracy. 

The mass discrepancy in intermediate- to high-mass eclipsing binaries is most commonly associated in the literature with deficiencies in stellar structure and evolution (SSE) models. In particular, multiple studies highlight poorly calibrated and/or missing input physics, such as interior mixing in SSE models as the cause of the mass discrepancy \citep[e.g.][and references therein]{Guinan2000,tkachenko_mass_2020}. However, owing to the method of inference of evolutionary masses, the mass discrepancy can also be regarded as a discrepancy between atmospheric (effective temperature, surface gravity) and fundamental (mass, radius) properties of stars. Indeed, evolutionary masses are inferred from fitting evolutionary tracks to the position of the star in the HR or Kiel diagrams. The mass discrepancy arises when the star appears hotter and/or more evolved than predicted by the evolutionary track concurrent with its dynamical mass \citep[e.g.][]{Pavlovski2009,Tkachenko2014}. In this context, the mass discrepancy phenomenon can no longer be isolated from models of stellar atmospheres and spectra, as those are used for the inference of the effective temperature, \teff, of the star. For example, a systematic overestimation of \teff\ would result in the mass discrepancy even if perfect SSE models existed and were used. Indeed, \citet{tkachenko_mass_2020} demonstrate that missing physics in models of stellar atmospheres may lead to an overestimation of \teff\ of about 8\% for an early B-type component of the V380 Cyg eclipsing binary system.

When thought of even more broadly, the question arises whether (often inhomogeneous or inconsistent) methodologies used for the analysis of individual systems can at least partially be the cause of the mass discrepancy phenomenon. In the study of double-lined eclipsing binaries, the analysis typically involves separate methods for photometric and spectroscopic observations, each optimised to handle distinct types of data. Despite these distinct roles, both photometric and spectroscopic analyses rely on a shared set of physical parameters that must be consistently optimised across datasets.
In traditional workflows, these shared parameters are often refined iteratively. Each observable is analysed separately, with parameters adjusted sequentially to achieve consistency. This iterative process may introduce biases and potential drifts across iterations. Such an approach can result in a non-optimal global solution, where parameter values diverge due to the interdependence of observables and the arbitrary starting values chosen. Moreover, an iterative approach is both time-consuming and computationally intensive, which limits scalability, making it impractical for studies requiring consistent analysis across large samples.

Last but not least, historically, the study of massive eclipsing binaries has often relied on an 'object-per-paper' approach, wherein individual systems are analysed independently by different research groups, each applying slightly different methodologies and assumptions. This fragmented approach has led to cumulative studies that, while informative, suffer from considerable scatter in the results due to variations in observational and analytical techniques. Consequently, no consistent, comprehensive observational map of the mass discrepancy exists across the parameter space. The issue is further compounded by the need for multi-epoch, high-resolution spectroscopic data, which requires extensive observational coverage to sample orbital phases for an accurate orbital solution. 

To summarise, when an overarching look is taken at the mass discrepancy problem and its possible causes, four distinct ingredients arise: (i) inhomogeneity of stellar samples used to study the phenomenon; (ii) methodological approach(es) employed for the analysis of individual binary systems; (iii) models of stellar atmospheres and spectra; and (iv) models of stellar structure and evolution. These four ingredients are listed in the order of the ease of testing and priority, as we see them. In our previous study \citep{tkachenko_observational_2024}, we started to explore the problem of stellar samples inhomogeneity by presenting an extensive sample of early spectral type TESS eclipsing binaries
for which we acquired phase-resolved high-resolution ($R\sim 85\,000$) spectroscopic time series as part of an ongoing large programme with the {\sc hermes} spectrograph \citep{Raskin2011} on the Mercator telescope. For the part of the sample observed to date (83 systems), we performed classification from both light curves and spectroscopy and obtained preliminary orbital solutions with the method of spectral disentangling \citep{Simon1994,Hadrava1995,ilijic_obtaining_2004}. 
This initial study laid the groundwork for testing and understanding the role of stellar samples inhomogeneity in the mass discrepancy problem. 

In this study, we take the next step and investigate the role of data analysis methodologies in the mass discrepancy problem. The limitations of traditional iterative methods for the analysis of individual binary systems underscore the need for a more integrated and efficient framework, particularly as we move toward larger-scale studies. To overcome these limitations, we introduce a novel framework that allows for a concurrent analysis of photometric and spectroscopic time-series data in a unified model. The central concept of this framework is to integrate light curve synthesis, spectral disentangling, and atmospheric parameters determination with spectral synthesis within a single model described by shared physical parameters. By simultaneously fitting the light curves and spectra, we ensure that the shared parameters are consistent across all datasets and get immediate feedback from light curve fitting, spectral disentangling, and spectral synthesis. 

In Section \ref{sec:classic}, we review traditional methods for the analysis of eclipsing binaries and discuss the limitations of iterative approaches. In Sections \ref{sec:framework}-\ref{sec:uncert}, we introduce our new framework, which integrates light curve and spectroscopic analysis into a unified model. Section \ref{sec:tests_syn} presents a series of tests on synthetic data to validate the framework’s ability to recover accurate stellar parameters. We applied this method to real binary systems, as described in Section \ref{sec:tests_obs}, illustrating its efficiency and robustness across a diverse mass range. Finally, in Section \ref{sec:conclusions}, we discuss the implications of our results for resolving the mass discrepancy problem and the potential for large-scale applications.

\section{Method}\label{sec:method}

The analysis of eclipsing binary systems typically involves iterative, separate analyses of light curves and spectra to extract fundamental stellar parameters. By observing both the photometric variations in the light curve and the spectroscopic features across orbital phases, we can determine a set of precise and accurate fundamental properties of stars such as mass, radius, and temperature, as well as the orbital parameters of the system \citep[e.g.][]{Torres2010}.

\subsection{Traditional approach}\label{sec:classic}

The light curve modelling for eclipsing binaries involves fitting observed brightness variations due to eclipses and orbital motion. High-precision light curves are extremely sensitive and are optimal for constraining the inclination, relative radii, and relative surface brightnesses of the stars.
A widely used tool for light curve fitting is PHysics Of Eclipsing BinariEs  \citep[{\sc phoebe},][]{prsa_computational_2005, prsa_physics_2016}, an advanced modelling package built upon the foundational Wilson-Devinney code \citep{Wilson-Devinney1971}.  {\sc phoebe} provides a detailed representation of binary systems by including multiple physically accurate effects such as heating and reflection, as well as surface distortion due to tidal forces. This enables the analysis of complex systems, including contact binaries; however,  {\sc phoebe}’s comprehensive approach requires significant computational resources. 
As an alternative, \texttt{ellc} \citep{maxted_ellc_2016} offers a faster and more efficient approach to light curve modelling. While it lacks some of the advanced physics present in  {\sc phoebe}, \texttt{ellc} is still accurate for most detached systems. Designed for efficiency, \texttt{ellc} models stars as triaxial ellipsoids, simulating their flux variations during eclipses with great speed and flexibility. This makes \texttt{ellc} a practical choice for applications requiring a high computational efficiency.

The analysis of spectra of binary stars is complicated by the fact that spectra observed at different orbital phases are actually composite signals, containing contributions from both stellar components shifted due to orbital motion. The resulting blending makes it difficult to extract individual atmospheric parameters like temperature, surface gravity, and chemical composition for each star. Measuring orbital phase-dependent shifts in terms of radial velocities (RVs) provides a means to  determine the orbital parameters of the system. In practice, obtaining accurate RV curves is often a challenging task, in particular at phases of the components' alignment along the line of sight and/or for systems composed of rapidly rotating stars with prominent rotational line broadening. Additionally, even with well-determined orbital parameters, the blended nature of composite spectra prevents direct access to the distinct spectral characteristics of each star. 

Spectral disentangling (SPD) is a technique that addresses the above-mentioned challenges by effectively separating the spectra of the individual components in a binary system. In SPD, individual 'single-star' spectra of each component star are treated as unknowns that, when shifted by the appropriate RVs and added up linearly, reproduce the observed time series of composite spectra. To recover these unknown spectra, we need to solve a matrix equation that accounts for the contributions of each shifted spectrum at the observed epochs. In addition to separating the spectra, SPD also optimises the orbital parameters, as they determine the RVs used in each instance of separation. The SPD technique was introduced by \citet{Simon1994}, and further expanded in Fourier space from wavelength space by \citet{Hadrava1995}. One of the most known implementations of spectral disentangling is fd3/FDBinary \citep{ilijic_obtaining_2004}, which performs the separation in the Fourier domain and uses the Nelder-Mead optimisation method to find the orbital parameters. As a result, the SPD technique returns the binary system's orbital elements and two individual spectra. The latter may further be used for inference of atmospheric parameters, employed as if the spectrum belongs to a single star with a signal-to-noise ratio (S/N) that has been enhanced (by effectively averaging over time). This approach is extremely convenient, although the quality of the resultant disentangled spectra suffers from multiple nuances, such as unknown light factors, edge effects, undulations, phase coverage, and so on. Therefore, a meticulous approach is required when using this technique.

When working with a double-lined eclipsing binary, we need to analyse its light curve and phase-resolved series of high-resolution spectra to have the most comprehensive parametrisation of the fundamental properties of both components. In practice, this implies an iterative approach to the analysis of different data types separately -- even if all of them originate from the same physical system and  could technically be parametrised by the same set of parameters. A non-exhaustive list of the parameters includes: 1) the masses that determine the orbital parameters of the system, which impact both the light curve modelling and SPD. The masses also contribute to \logg, one of the key parameters in the calculation of atmosphere models and synthetic spectra;  2) the effective temperatures that determine the morphology of the spectrum and also determine the flux ratio parameter used in the light curve modelling; 3)  limb darkening laws are foundational for accurate modelling of the eclipse shapes in light curves and are determined by the same atmosphere models that are used to compute synthetic spectra; 4) light factors in SPD basically represent individual flux contributions of the components to the light curve of the system. 
In an iterative approach, one has to deal with one subset of parameters at a time while fixing the rest at arbitrary initial values. The latter are often inconsistent with observables sensitive to the corresponding parameters, an issue that is supposed to be resolved in subsequent iterations. An exchange of information between different sets of observables and free parameters continues until full convergence, which is often difficult to achieve in a self-consistent manner. The iterative approach is therefore expected to introduce an unquantifiable bias in the search for a global solution. 

\subsection{The new framework for simultaneous photometric and spectroscopic analysis}\label{sec:framework}

\begin{figure*}
   \centering
\includegraphics[width=170mm]{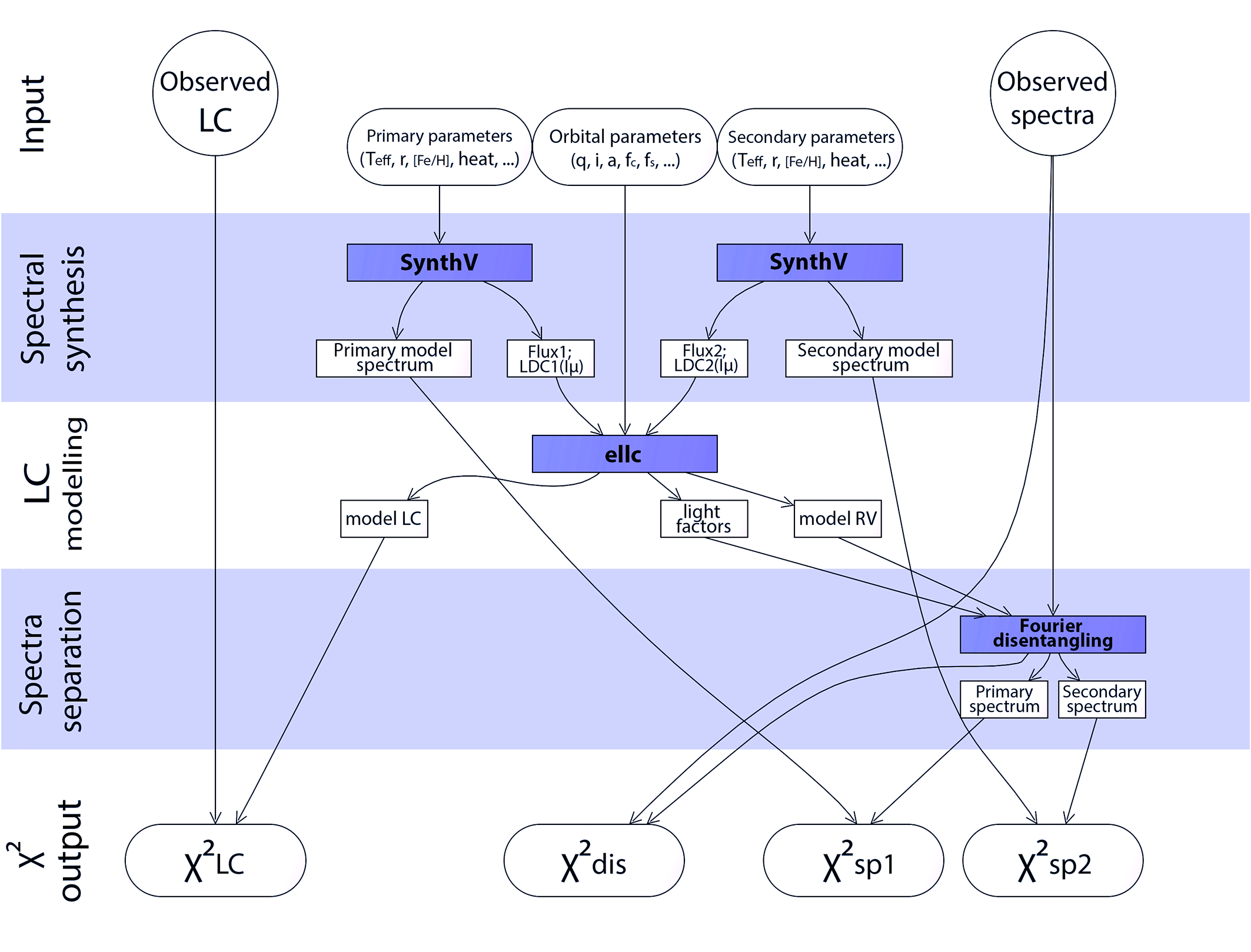}  
      \caption{Schematic presentation of the self-consistent model of photometric and spectroscopic time-series of EB.}
    \label{Fig:repairScheme}
\end{figure*}

The central concept of a new analysis framework introduced in this paper concerns generating a model capable of describing all the observed data for a given system. In practice, this implies merging the methods for the analysis of light curves and spectroscopic time series into a self-contained approach. Ideally, we would opt for completely redefining existing models for computing synthetic light curves and spectra based on a single model handling radiation transfer in the individual atmospheres of both components of the system, taking into account phase-dependent geometry and surface brightness variations to compute time series of the composite spectra, as well as the integrated total flux of the system. However, this maximally correct approach would require immense computational resources and make the process of optimisation (search for optimal parameters for a given binary system) an impossible task. At the same time, the current commonly used approach of juggling different models for spectroscopic and photometric data of one system with iterative refinement of cross-dependent parameters is by nature subjective and plagued by biases and accumulating errors that are hard (if not impossible) to quantify. 

In this work, we propose a new approach that is a trade-off between the above-presented extremes. We exploited ready-to-use solutions for on-the-fly spectral and light curve synthesis, simplifying the model spectra computation by working with orbit-averaged separated spectral components of the two stars and re-using the information on synthesised non-normalised intensity spectra to inform the light-curve modelling tool on the atmospheric properties of the components.  

A schematic overview of the model generation inside each optimisation iteration is depicted in Fig. \ref{Fig:repairScheme} for the case of an SB2 system with available orbital phase-resolved time series of composite spectra. The system may be described by one set of parameters that characterise its orbit [$i$, $a$, $f_c$, $f_s$, $P_{orb}$, $T_0$, ...] and the fundamental and atmospheric parameters of both components [$r$, \teff, $V_e$, [M/H], ... ]. Here, $i$ is the orbital inclination, $a$ is the semi-major axis, $f_c$ and $f_s$ are reparametrisations of the eccentricity, $e$, and argument of periastron, $\omega$, as $f_c \equiv \sqrt{e} \cos(\omega)$, $f_s \equiv \sqrt{e} \sin(\omega)$, $P_{orb}$ is the orbital period, $T_0$ is the reference time of periastron passage, $r$ is the radius relative to the semi-major axis, \teff\ is the effective temperature, $V_e$ is the equatorial rotational velocity, and [M/H] is the metallicity. These parameters, together with the observed light curve and spectra, serve as input and are denoted in the corresponding 'input' block of  Fig. \ref{Fig:repairScheme}. 

First, we computed individual model spectra of the two components of a binary. In the simplest case, we need [\teff, \logg, \vsini, [M/H]] to compute a synthetic spectrum, which can be directly obtained from subset [\teff, $r$, $a$, $V_e$, $i$, [M/H]] of the full set of parameters of the system (while the absolute masses, $M,$ can be computed from orbital parameters). The resultant spectra (including specific intensities $I_{\mu}$, where $\mu=\cos\theta$ defines the position on the stellar disk) allow us to (i) compute the brightness ratio of the components (integrated over the photometric passband) and pass it to the light curve modelling tool along with $I_{\mu}$ for each component to allow for an accurate computation of the limb-darkening effect, instead of an ad hoc law with free parameters; (ii) use these theoretical spectra for a comparison with the disentangled ones. The spectra need to be computed in both wavelength ranges of the photometric passband and the observed spectra. This step is denoted as the 'spectral synthesis block' in Fig. \ref{Fig:repairScheme}. The {\sc SynthV} \citep{tsymbal_starsp_1996, tkachenko_grid_2015} code was employed in this step for the LTE spectral synthesis, which requires a precomputed atmosphere model as input. We used pre-computed grid of models {\sc LLModels} \citep{kochukhov_stellar_2005,khan_stellar_2006,khan_stellar_2006-1} and use cubic linear interpolation to interpolate the grid nodes onto the requested point in  [\teff, \logg, [M/H]]; however, we note that any grid of Kurucz's original format \citep{kurucz_model_1979} would be generally supported.

In the next step, we computed a model light curve based on the orbital parameters [$i$, $a$, $f_c$, $f_s$, $P_{orb}$, $T_0$], the fundamental properties of the individual components [$M_1$, $M_1$/$M_2$, $R_1$, $R_2$, $I_{\mu 1}$, $I_{\mu 2}$ ], and the light curve-specific coefficients such as heating, gravity darkening, third light, and so on. This step is denoted as the LC modelling block in Fig. \ref{Fig:repairScheme}. Apart from the light curve itself, this step produces an RV curve for a given set of orbital parameters and the individual component fluxes that allow us to infer phase-dependent light factors for the next step. For this step, we ran the \texttt{ellc} code by \citet{maxted_ellc_2016}.

In the next step, we performed a spectral separation to reconstruct individual spectra of the two binary components from the observed series of composite spectra based on the RV curve and light factors computed for the epochs of observation in the previous step. This step produces two 'observed' phase-averaged individual spectra of binary components, which is denoted as the 'spectra separation' block in Fig. \ref{Fig:repairScheme}. This step is performed with to Python module adapted from {\sc fd3} \citep{ilijic_fd3_2017} with Fourier spectral separation. 

After computing the model data, the goodness-of-fit is assessed using a modified statistic based on normalised residuals, defined as:

\begin{equation}
\chi^2 = \sum_i \frac{\left(F_i - M_i\right)^2}{M_i^2},
\end{equation}

where $F_i$ represents the observed data and $M_i$ is the corresponding model prediction. This formulation, commonly referred to as a relative residual sum of squares (RRSS), provides a practical alternative to the traditional $\chi^2$ statistic, which uses observational uncertainties, $\sigma_i$, in the denominator. In this work, $\sigma_i$ is excluded for several reasons. For disentangled spectra, which constitute two out of the four $\chi^2$ terms, observational uncertainties are not well-defined. The disentangled spectra are derived products that depend on the orbital solution, meaning their quality (and thus their uncertainties) evolves dynamically throughout the optimisation process. Using a dynamic $\sigma_i$ tied to the optimisation state would be statistically inconsistent. For period-folded light curves, uncertainties are influenced by data de-trending and phase-folding. Additionally, the light curves are rebinned non-uniformly to over-sample the eclipse phases, further complicating the definition of reliable $\sigma_i$ values. The RRSS approach ensures a stable, reproducible measurement of the goodness-of-fit across all datasets. 
The modified $\chi^2$ statistic is computed for four components, each tailored to specific observables:

\begin{itemize}
    \item \textbf{$\chi^2_{LC}$}: This term evaluates the fit between the observed light curve ($F_{LC}$) and the model light curve ($M_{LC}$). The model light curve is computed by integrating the specific intensities $I_{\mu}$ of synthetic spectra over the observational passband and using these $I_{\mu}$ as input to the \texttt{ellc} light curve synthesis module as surface brightness ratio and limb darkening law. The observed light curve is a period-folded, rebinned dataset, where binning is non-uniform to over-sample the eclipse phases. This $\chi^2$ depends on the system's orbital parameters (e.g. inclination, eccentricity, and semi-major axis), the fundamental parameters of the binary components (e.g. $T_{\rm eff}$, $\log g$), and light curve-specific parameters such as heating, gravity darkening coefficients, and so on.
    
    \item \textbf{$\chi^2_{Dis}$}: This term compares the observed composite spectra ($F_{Dis}$) with the reconstructed composite spectra ($M_{Dis}$). The reconstructed composite spectra are generated as a linear sum of the separated spectra for the primary and secondary components, each shifted according to their predicted radial velocities (RVs) from the current orbital solution. The light factors, derived from the light curve relative fluxes, act as proxies for the radii ratio and surface brightness variations. $\chi^2_{Dis}$ is highly sensitive to the orbital parameters and phase-dependent light factors, making it a critical term for constraining the orbital solution.
    
    \item \textbf{$\chi^2_{Sp1}$}: This term assesses the agreement between the primary's separated spectrum ($F_{Sp1}$) and the synthetic model spectrum ($M_{Sp1}$). The synthetic spectrum is computed using the primary's fundamental parameters, including $T_{\rm eff}$, $\log g$, metallicity, and rotational broadening. The 'observed' separated spectrum is obtained through the disentangling process and depends on the quality of the orbital solution. As a result, this $\chi^2$ is sensitive to both the accuracy of the orbital solution and the fundamental parameters of the primary component.
    
    \item \textbf{$\chi^2_{Sp2}$}: Similarly to $\chi^2_{Sp1}$, this term compares the secondary's separated spectrum ($F_{Sp2}$) with its synthetic model spectrum ($M_{Sp2}$). The model spectrum depends on the secondary's fundamental parameters and the 'observed' spectrum is influenced by the same disentangling assumptions as the primary.
\end{itemize}

By minimising all four $\chi^2$ terms simultaneously -- either as a weighted sum or in a multi-objective optimisation framework -- the model can consistently refine the full set of system parameters. This approach avoids having to make iterative adjustments between different observables and exploits the interconnected nature of light curves and spectral modelling. 
Each of these $\chi^2$ values contains information about goodness-of-fit for the full unified set of parameters common for all the steps. $\chi^2_{LC}$ depends on the system orbital parameters, fundamental parameters of the components, and a few light curve-specific free parameters; $\chi^2_{Dis}$ depends mostly on the system orbital parameters (via RVs) and phase-dependent light-factors determined from the light curve fitting (which act as a proxy of radii ratio and surface brightness variations); $\chi^2_{Sp1}$ and $\chi^2_{Sp2}$ depend on the fundamental parameters of individual binary components and quality of the spectral separation. In this way, all observed data provide feedback for one full set of parameters in a self-contained manner.  Minimising all four $\chi^2$ terms simultaneously (either as a weighted sum of all four or in a multi-objective optimisation) allows us to reconstruct the full set of parameters of the system, while avoiding having to perform iterations between different types of observables.

The advantages of such an approach are: i) a single, self-consistent orbital configuration is used to model the light curve and perform the separation of the components' spectra; ii) no add-hoc limb-darkening law is required, since the radiative transfer used for spectral synthesis provides a set of $I_{\mu}$ values to model the effect accurately in a given photometric passband; iii) using one collapsed set of parameters to create a model for different observations simultaneously allows to limit the level of underdetermination while fitting; and iv) the disentangling may be performed on a large wavelength range, in contrast to a few-lines-at-a-time approach in stand-alone disentangling. In the unified framework, not only can we use the model spectrum to 'iron out' any
undulations, but also the light factors do not need to be fixed at the generic value of 0.5. The latter requires further spectra rescaling, together with undulations. In this work, we have computed the individual flux contributions from light curve modelling, which are orbit-variable consistently with light cures, which improve the stability of spectra separation.

A number of challenges still exist in this approach and should be acknowledged: i) limited classes of systems may be modelled as the parameter range is constrained by the atmosphere models -- both components should not be compact or supergiants (as the grids are limited in \logg); ii) the use of \texttt{ellc} tool for limits applications further to detached systems; iii) no orbital variability is accounted for in the model spectra, as the disentangled spectrum is assumed to be orbit-averaged (= full orbit is uniformly sampled by the spectral time-series) and may be represented by a model spectrum of a normal single star, while this is not guaranteed for systems with extreme heating and/or distortion; and iv)  on-the-fly spectra require extended computation time. This is especially pronounced with wide spectral ranges. Thus, we are limited to short ranges, which we describe below. In the following sections, we use a range of 4250-4500\AA\  to cover one hydrogen line and important helium and metal lines. This takes four seconds on one core out of roughly four seconds needed for the entire model, including spectra separation and light curve model. The wide TESS photometric passband needs to be covered fully, but there we can compute model with no metal lines, so this part is negligible time-wise.

\subsection{Multi-objective optimisation}\label{sec:multi-objective}

\begin{figure}[h!]
   \centering
\includegraphics[clip,width=80mm,trim={0.5cm 1cm 1cm 0cm}]{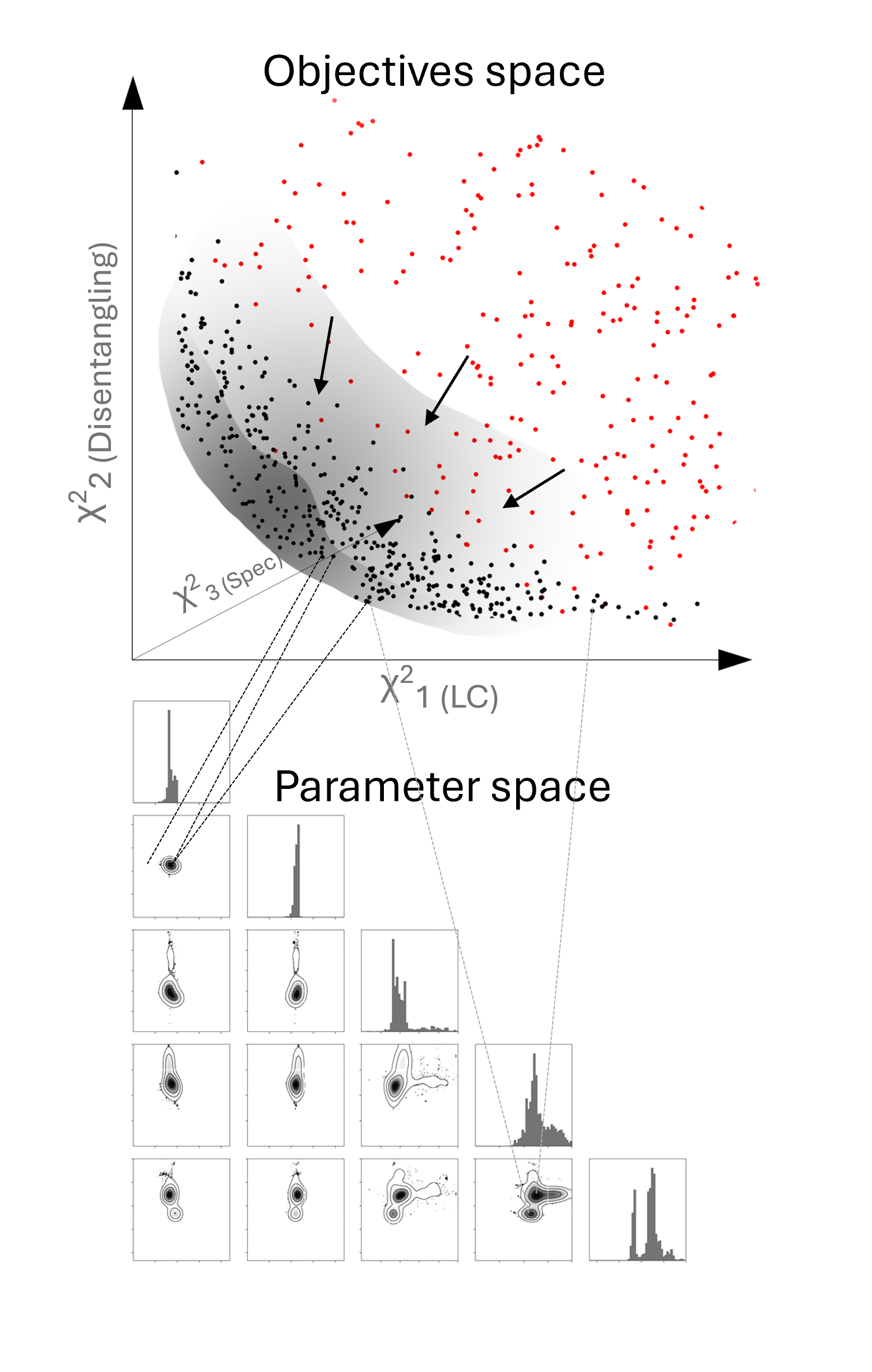}  
      \caption{Upper panel:\ Visual representation of a cloud of solutions (sets of parameters) in 3D space of objective functions, advancing from initial population state (red) towards approaching the trade-off surface (black). Lower panel:\ Representation in the parameter space via pairwise corner plot. Solutions that are close in the objective space (i.e. have similar $\chi^2$ values) do not have to appear close in the parameter space, while those close in the full parameter space should appear close in the objectives space if the parametrisation is continuous.  }
    \label{Fig:mooScheme}
\end{figure}

As described in the previous section, the new framework for modelling double-lined eclipsing binary systems requires simultaneous optimisation of four objective functions, each one representing a different aspect of the observational data: light curve fitting ($\chi^2_{LC}$), spectral disentangling ($\chi^2_{Dis}$), and fitting the disentangled spectra of the primary ($\chi^2_{Sp1}$) and secondary ($\chi^2_{Sp2}$) components. A common approach to handling such a problem would either be to fit the different objectives iteratively, or assign weights to the different objectives and minimise a weighted sum of the objective functions.
However, these approaches introduce several complications. In the former case, iterative fitting can lead to conflicting updates between different steps. In the latter approach, the assignment of weights is often rather arbitrary and depends on subjective choices that may bias the results. Instead, we can resort to a more powerful class of techniques known as multi-objective optimisation.

Multi-objective optimisation (MOO) is a class of optimisation methods specifically designed to handle problems where several conflicting objectives must be optimised simultaneously. Unlike traditional single-objective optimisation, which seeks to minimise (or maximise) a single function, MOO aims to find a surface of solutions in N-dimensional space of objectives \citep{srinivas_muiltiobjective_1994}. In MOO, there is no single 'best' solution. Instead, the algorithm searches for a set of solutions from a trade-off surface in the space of objective functions (see a schematic illustration of the MOO trade-off surface in Fig. \ref{Fig:mooScheme}). Each point on this surface corresponds to a solution where improving one objective can only come at the cost of worsening one  other objective (or even several). Without defining weights, solutions from this asymptotic trade-off surface are by definition equal in terms of the overall cost for a set of all objective functions.
For the current framework, this means that instead of prioritising the fit of one observable at the time (e.g. light curve) over the others, MOO enables us to explore a set of solutions that provide a balanced fit across all four objectives: light curve fitting $\chi^2_{LC}$,  spectral disentangling $\chi^2_{Dis}$, and fitting of the primary's $\chi^2_{Sp1}$ and secondary's $\chi^2_{Sp2}$ disentangled spectra.

To perform the multi-objective optimisation, we employed a widely used algorithm called Non-dominated Sorting Genetic Algorithm II (NSGA-II). NSGA-II \citep{deb_fast_2002} is an evolutionary algorithm that works by evolving a population of candidate solutions over multiple generations. It is particularly well-suited to MOO problems because it effectively searches the space of objectives, identifying the trade-off surface where the solutions lie.
Central to NSGA-II is the concept of Pareto dominance \citep{deb_fast_2002}. In MOO, a solution \( \mathbf{x}_1 \) dominates over another solution \( \mathbf{x}_2 \) if it is at least as good in all objectives and strictly better in at least one:
\begin{align}
    \forall i \in \{1, \dots, m\}, \, f_i(\mathbf{x}_1) &\leq f_i(\mathbf{x}_2) \notag \\
    \text{and} \quad \exists j \in \{1, \dots, m\} \, &\text{such that} \, f_j(\mathbf{x}_1) < f_j(\mathbf{x}_2) .
\end{align}
A solution is said to be optimal if it is not dominated by any other solution in the population. The set of all optimal solutions forms the Pareto front, representing the best trade-offs between the objectives. Solutions on the Pareto front are non-dominated, meaning no other solution is strictly better in all objectives. The NSGA-II algorithm identifies solutions along the Pareto front by:
\begin{enumerate}
    \item Non-dominated sorting: Each solution is assigned to a Pareto front based on how many solutions dominate it. The first front consists of solutions that are not dominated by any other, while subsequent fronts contain solutions that are dominated only by solutions from the previous fronts but dominate the rest solutions.
    \item Crowding distance: To maintain diversity among solutions and avoid clustering, NSGA-II calculates a crowding distance for each solution. Solutions that are farther from others in the objective space are more likely to be selected for the next generation.
    \item Evolutionary operators: The algorithm evolves the population through operations such as selection, crossover, and mutation, iteratively refining the solutions over several generations. As a result, the population approaches the true optimal Pareto front.
\end{enumerate}

A number of different packages with NSGA-II implementation exist ready-to-use. We opted to use DEAP Python library \citep{fortin_deap_2012}. DEAP provides a user-friendly yet highly customizable framework for multi-objective optimisation.
By employing NSGA-II, we can efficiently search the parameter space of the binary system and find a set of solutions that provide a balanced fit to the light curve and spectroscopic data. Each solution on the Pareto front represents a valid trade-off between fitting the light curve, disentangling the spectra, and fitting the individual stellar spectra. 

In our case, the parameters of the binary system \( \mathbf{x} \) (e.g. orbital parameters, stellar radii, temperatures, and surface gravity) were simultaneously optimised across all four objectives. NSGA-II ensures that we do not need to assign arbitrary weights to prioritise one objective over another, allowing us to explore a diverse set of solutions that respect all the available observational constraints. This multi-objective approach gives us a flexible and robust way to model multi-dataset observations, minimising the inherent biases and inconsistencies that arise from iterative or weighted optimisation methods. Moreover, this approach (as with genetic algorithms in general) does not require an initial guess: the initial population is created as a Sobol grid, covering the full range of parameter space of interest.

\subsection{Computational considerations}\label{sec:computations}

The most computationally intensive part of our framework is the on-the-fly calculation of synthetic spectra during model evaluations. For each star in the binary, the spectral synthesis is required not only to obtain the disk-integrated flux for the spectrum model, but also to obtain specific intensity profiles $I(\mu)$ across the photometric passband to inform light curve model on light ratio and limb darkening consistently with spectra of both stars. To optimise performance, line opacities can be neglected for the broad photometric bands, using only continuum, hydrogen, and helium contributions, which considerably reduces the computational time. Nevertheless, depending on the spectral resolution and interval size, each spectral synthesis can take from 2 to 20 seconds, dominating the computational cost per model. Other components of the model, including  atmospheric model interpolations, light curve synthesis, and spectral separation, require comparatively negligible time (milliseconds per model). In further tests, we opted for using a short spectral interval of only 100\AA~ for the high-resolution line spectrum, resulting in 4 seconds per full model computations (two spectra, $I(\mu)$ in photometric passband, light curve of the system, spectral separation, and evaluating residuals for the full spectroscopic and photometric time series).  

Since the genetic algorithm requires evaluating a large number of models (and especially so in the case of MOO) across a broad parameter space, the total number of model evaluations can reach tens of thousands to a million, depending on the dimensionality of the problem and the search volume. Typically, between 7 and 15 free parameters are optimised for one system, and the number of evaluations scales with the product of the population size and the number of generations. However, the genetic algorithm is naturally parallelisable, with different individuals evaluated independently. This allows for efficient scaling across multiple cores. In practice, analysing a single binary system typically requires 1 to 2 days (depending on the number of free parameters) on a system with 20 cores. 

Given these computational characteristics, our framework is particularly suited for intermediate-sized samples. Very small samples can be more efficiently treated by traditional iterative methods with full manual control over detailed physical effects and nuances of each individual system, while very large samples would require heavy computational resources. The intended primary application of this framework is the analysis of a sample containing approximately 100 systems in the next paper in this series, which remains computationally feasible with currently available resources.

Future improvements could include precomputing grids of $I(\mu)$ for photometric passbands and coupling them to spectra in precomputed grids, which would both lift the most time-demanding computation step and relieve the solid dependency on the single spectral synthesis code. Additionally, experiments with optimisation schemes that require us to compute objectives fewer times are possible, although it is challenging for effective parameter space exploration during MOO. Establishing a   dynamic scheme for searching weights for collapsing several objectives of MOO into a single objective could help in reducing the complexity of the optimisation.  

\subsection{Uncertainty estimation for derived parameters}\label{sec:uncert}

The uncertainties of the derived parameters are computed based on the statistical properties of the last generations of the evolved population obtained in the MOO framework. Specifically, we estimate uncertainties from the confidence ellipses of the last generations of the GA. This approach provides a clear visual and numerical representation of the parameter correlations and allows us to extract a simple interpretable measure of uncertainty, such as the 1$\sigma$ confidence levels. An example of these confidence ellipses is shown in Fig. \ref{Fig:B5_conf_ell} on top of density levels in the corner plot of final generations. We can see that while capturing correlations of parameters in the case of a single minimum, the confidence ellipses struggle with double or multiple minima and show large common uncertainty rather than uncertainty per minimum.  In cases where local minima are clearly distinguishable, we may either filter secondary minima out, or resort to deriving uncertainties from the density levels in the same corner plots, which better account for more complex or irregularly shaped distributions.  In cases with complex structure of minima, the parameter uncertainties should either be reported for each minimum separately or treated as upper limits encompassing the spread between minima. 

As an alternative, uncertainties can be estimated through local single-objective optimisation starting from the initial guess found in MOO, with a gradient-based method used to minimise a single collapsed objective function.  While this method offers a theoretically robust way to compute uncertainties, it has several practical limitations. First, the results depend on the way residuals of the four observables are combined into a single vector and which weights were used in this process, which introduces subjective choices and potential biases. Second, the method is highly sensitive to the choice of free and fixed parameters, as well as the complexity of the collapsed objective's landscape. In practice, local optimisation often struggles to converge in the collapsed with arbitrary weights landscape. Nonetheless, we have included this option in our framework, but it is much less autonomous and requires input weights.

Given these considerations, we adopted the confidence ellipse approach as the main for its balance of simplicity, interpretability, and robustness. Nonetheless, the potential for multi-minima scenarios and the limitations of this method should be acknowledged, and a rigorous look at all available information obtained in MOO should be considered before reporting any uncertainty exported by our framework.

\section{Tests on synthetic data}\label{sec:tests_syn}

\begin{figure*}
   \centering
\includegraphics[clip,width=180mm,trim={0.2cm 0.2cm 0.2cm 0.2cm}]{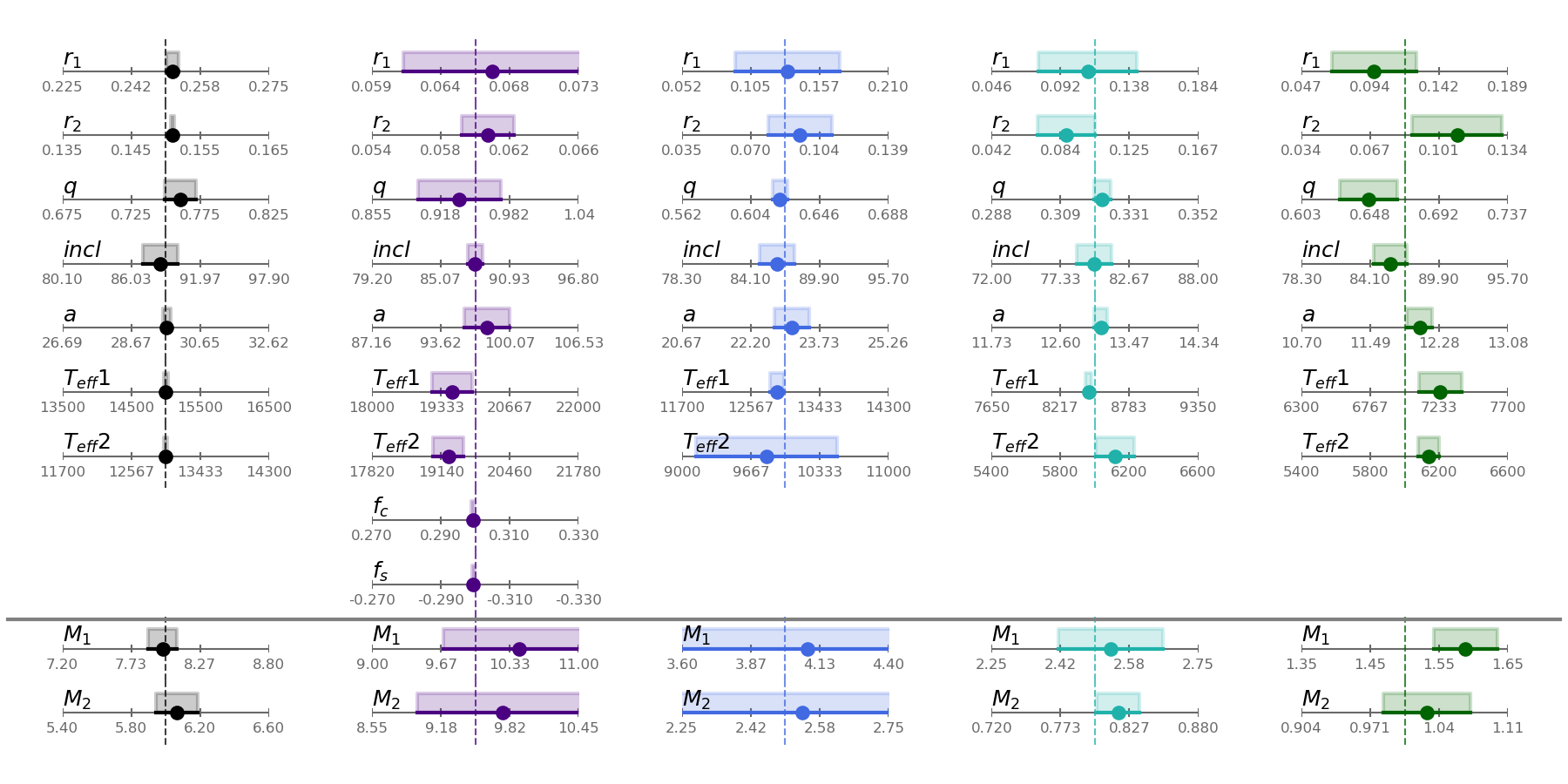}  
      \caption{Results of the tests on model data presented as deviations of the found solution from true parameters of the model. Left to right: Model systems referred to in the text and Table \ref{tab:synthetic_systems}, from B5+B7 to F2+G0. The dashed vertical lines show the true value of each optimised parameter. The value found by the optimiser and the associated 1$\sigma$ confidence interval are shown with the filled circle and shaded area, respectively. Each axis has a range of $\pm$10\% of true value so that the deviations may be visually compared between parameters and systems. An exception is made for the $r_1$ and $r_2$ pair of parameters for the last three systems, where axes had to be extended to accommodate large error bars (the numerical axes ticks should be used as guidance for these systems). Note: the masses of components are not optimised, they are computed from the optimised orbital parameters.}
    \label{Fig:synthsres}
\end{figure*}

\begin{figure*}
   \centering
\includegraphics[clip,width=175mm,trim={0.2cm 0.2cm 0.2cm 0.2cm}]{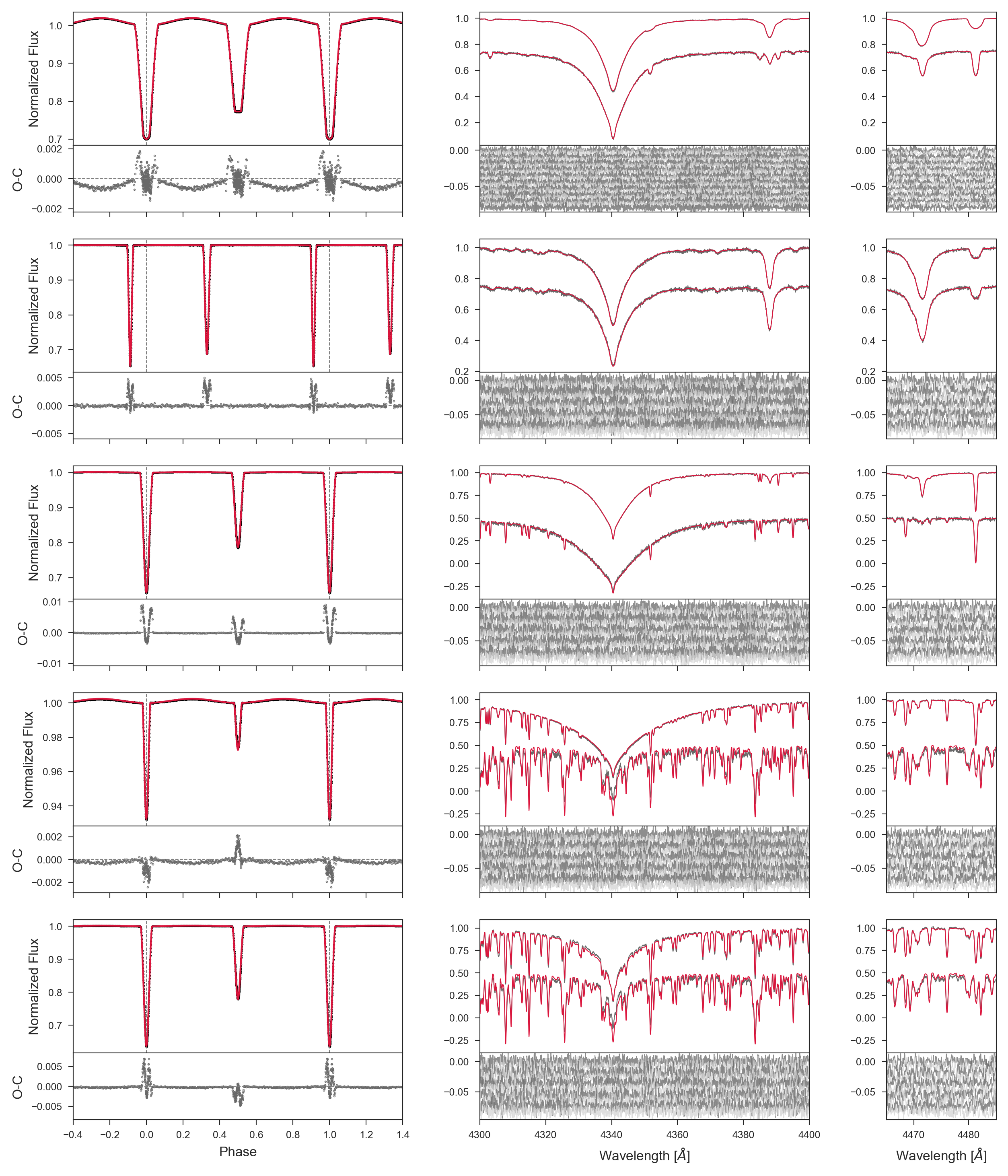}  
      \caption{Optimised solutions for all five artificial binary systems. From top to bottom: Model systems as referred in the text and Table \ref{tab:synthetic_systems}, from B5+B7 to F2+G0. Left column:  Binned model (red) versus noised and binned 'observed' (dark grey) light curves with the residuals beneath. Note: residuals inside eclipses appear more noisy due to eclipse over-sampling during the binning process. Middle and right columns: Disentangled spectra of both components (dark grey) and best fit synthetic spectra (red) in two wavelength regions; there is a wider range with the $H_\gamma$ line (middle) and a zoom into the region with the magnesium 4481\AA\ and the helium 4471\AA\ (for hotter stars) spectral lines (right). Disentangled spectra of the primary and secondary are shifted vertically for better visibility, residuals are shown beneath.}
    \label{Fig:synthsmulti_merged}
\end{figure*}

\begin{table*}
\caption{Configurations of the five synthetic test systems.}
\centering
\label{tab:synthetic_systems}
\begin{tabular}{cccccccccccc}

\textbf{System} & \textbf{M$_1$ [M$_\odot$]} & $q$ & $r_1$ & $r_2$ & $T_{\text{eff1}}$ [K] & $T_{\text{eff2}}$ [K] & $i$ [degrees] & $a$ [$R_\odot$] & $f_c$ & $f_s$ & P [days] \\ 
\hline
\hline
\textbf{B5+B7} & \textbf{8.0} & 0.75 & 0.25 & 0.15 & 15000 & 13000 &  89.0 & 29.657 & 0.0 & 0.0 & 5.0 \\
\hline
\textbf{B2+B2} & \textbf{10.0} & 0.95 & 0.066 & 0.060 & 20000 & 19800 &  88.0 & 96.845 & 0.3 & -0.3 & 25.0 \\
\textbf{B7+B9} & \textbf{4.0} & 0.625  & 0.131 & 0.087 & 13000 & 10000 & 87.0 & 22.9644 & 0.0 & 0.0 & 5.0 \\
\textbf{A4+G0} & \textbf{2.5} & 0.32 & 0.115 & 0.1045 & 8500 & 6000 & 80.0 & 13.033 & 0.0 & 0.0 & 3.0 \\
\textbf{F2+G0} & \textbf{1.5} & 0.67 & 0.118 & 0.084 & 7000 & 6000 & 87.0 & 11.888 & 0.0 & 0.0 & 3.0 \\

\end{tabular}
\end{table*}

\begin{figure*}
   \centering
\includegraphics[clip,width=190mm,trim={3cm 0cm 2cm 0cm}]{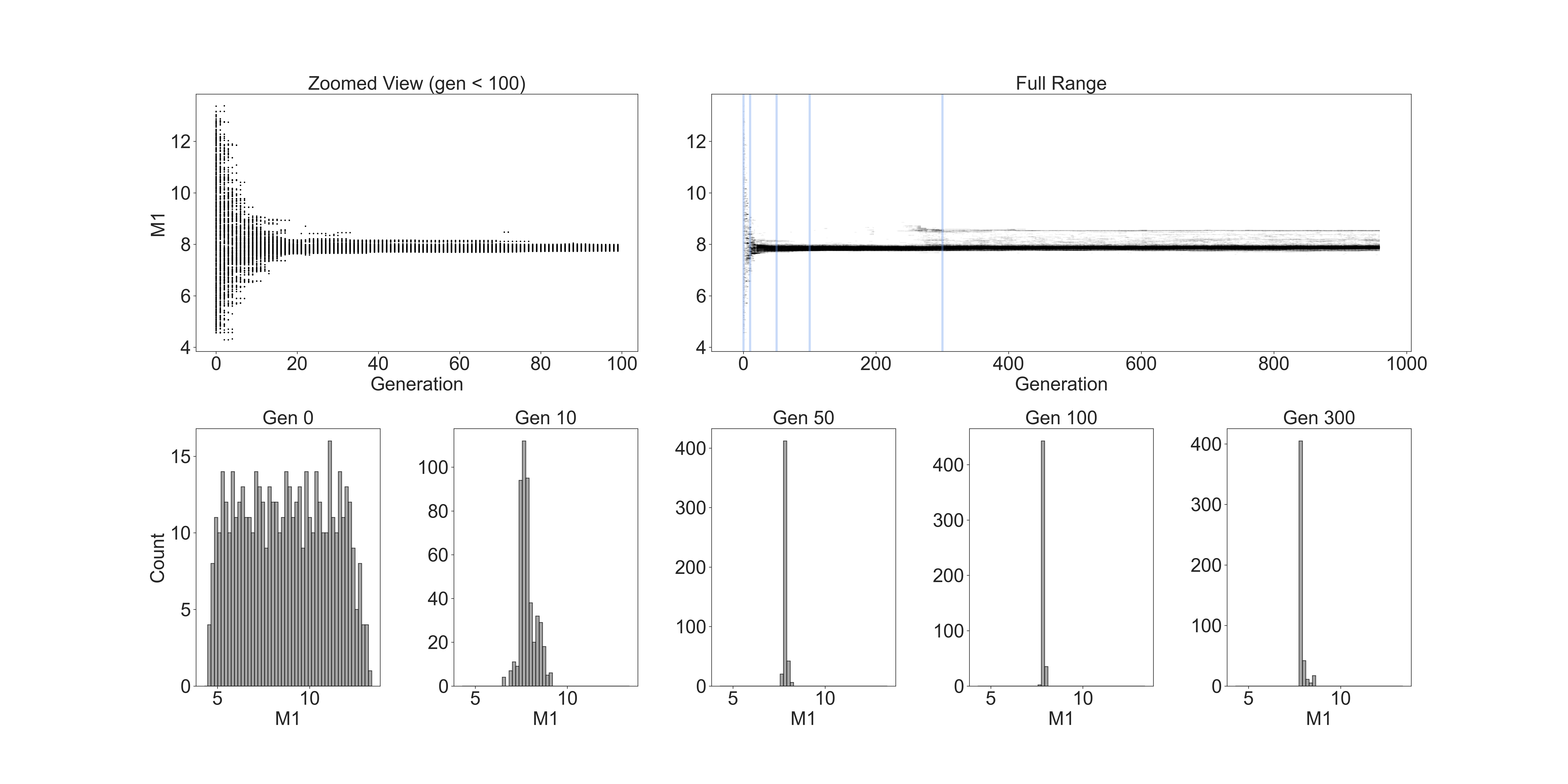}  
      \caption{Evolution of the primary's mass distribution along the population during optimisation for the B5+B7 artificial system. Top row: Value of the primary's mass for a single individual in the population, with populations evolving as the generation number increases. The left and right panels show a zoom into the first 100 generations and the full evolution, respectively. Blue vertical lines in the right panel indicate the positions of the slices shown in the bottom panels. Bottom row: Slices in the form of histograms for generations 0, 10, 50, 100, and 300 indicated with blue vertical lines in the top right panel.}
    \label{Fig:B5_evoM1}
\end{figure*}

\begin{figure*}
   \centering
\includegraphics[clip,width=190mm,trim={3cm 0cm 2cm 0cm}]{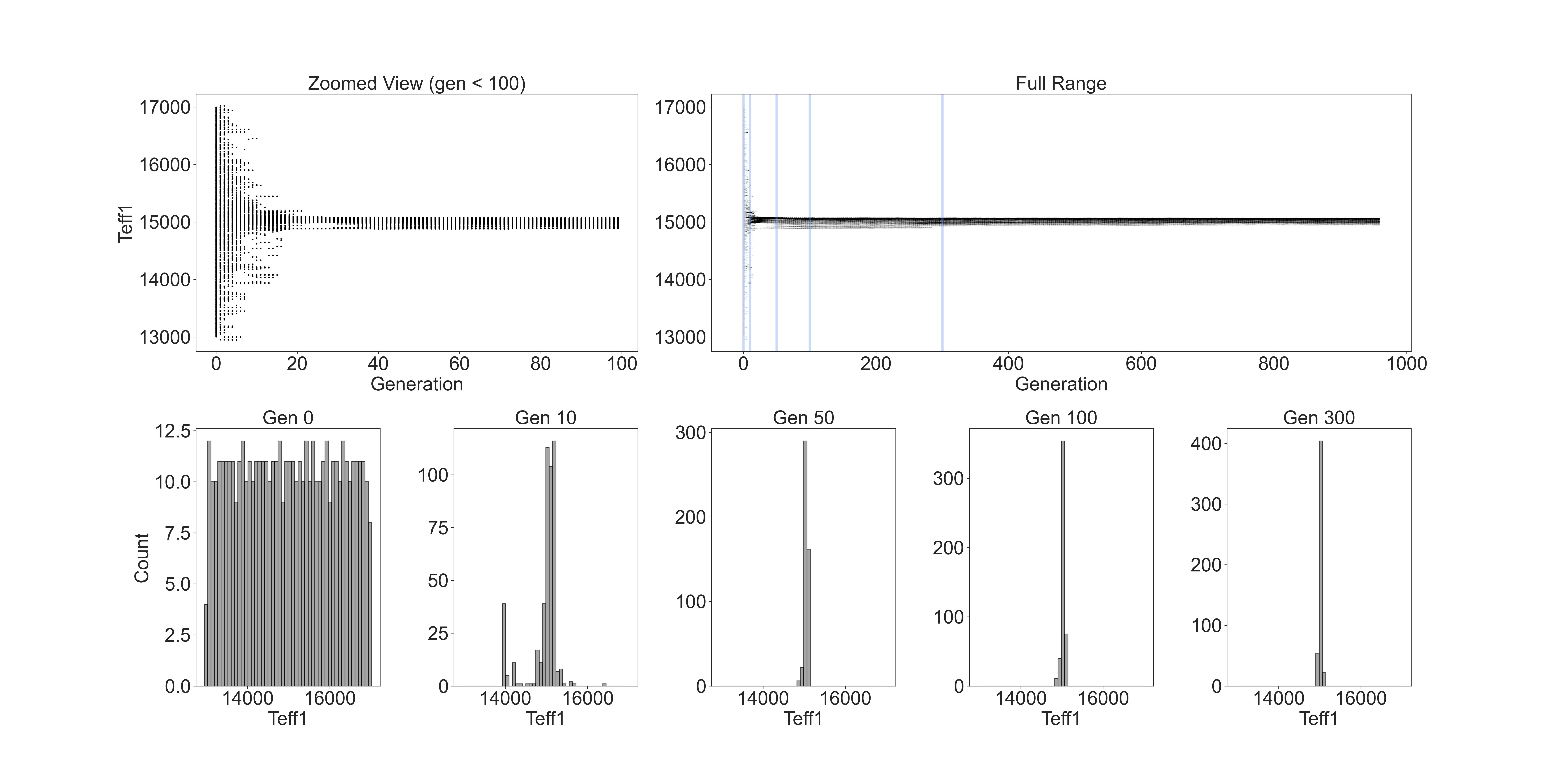}  
      \caption{ Same as Fig. \ref{Fig:B5_evoM1}, but for \teff.  }
    \label{Fig:B5_evoTeff}
\end{figure*}

To validate our modelling framework, we conducted tests using synthetic data. The tests involved generating synthetic light curves and composite spectra (both with added noise) for five model systems and running them through our framework. The goal was to assess whether the framework could accurately recover the original system parameters, including stellar radii, temperatures, and orbital characteristics. 

The synthetic data was generated using known parameters representative of typical eclipsing binary systems and their values were selected to span a range of stellar and orbital configurations (aiming to cover different temperature and mass regimes, light ratios, and mass ratios). The parameters of each system are listed in Table \ref{tab:synthetic_systems}, where they are named according to spectral types of the components: B5+B7, B2+B2, B7+B9, A4+G0, and F2+G0. Realistic white noise levels were added to simulate observational conditions. The noise was applied to both the light curves (S/N = 1000 for most cases, except for B2+B2, where we tested lower S/N of 200) and spectra (S/N = 150). The number of spectra covering the orbital period was set as ten, with uniform space coverage to recreate conditions similar to intended use on our full sample of the {\sc hermes} massive EB campaign \citep{tkachenko_observational_2024}. The only exception is the first system, B5+B7, which we created to be a perfect-case scenario with 20 spectra covering the orbital period, having full eclipses, no eccentricity, nearly equally bright components, and spectral lines easily distinguishable by different rotational broadening, to see whether MOO is capable of finding a single exact solution in the perfect case. The rest four systems were designed to have one of the following realistic complications: partial eclipses, eccentricity, low surface brightness ratio, heavily blended lines, etc.

Each of the five test systems was run through the full framework, which included global multi-objective optimisation. Each model computation included the separation of composite spectra and computation of the model light curve and synthetic spectra. The minimal set of parameters was optimised during the test: [$T_{\text{eff1}}$, $T_{\text{eff2}}$, $r_1$, $r_2$, $q$, $i$, $a$], as well as $f_c$ and $f_s$ (see Section~\ref{sec:framework} for definitions) for the first system that is eccentric. The full configuration file for the first system is provided in \href{https://zenodo.org/records/15388159}{Appendix D}. The fitting accuracy was assessed by examining the deviations of the derived parameters from the true values, as illustrated in Fig. \ref{Fig:synthsres}. Uncertainties were quantified using the final few populations of the parameter space (presented in the corner plot in Figs. \href{https://zenodo.org/records/15388159}{E.1-E.5}). The light curve fits and both components' spectra fits are presented in Fig. \ref{Fig:synthsmulti_merged} for all five systems. 

For the first ideal-case system (left column in Fig.~\ref{Fig:synthsres}), the framework found the exact correct parameters. Most solutions in the last population are concentrated in a compact circle with $\sigma$ of 1-2\%, depending on the parameter (see corner-plots in \href{https://zenodo.org/records/15388159}{Fig. E.5}. The orbital inclination is a notable exception, where the population is spread out in a much wider range. We interpret this finding as the orbital inclination being the most perceptive of the trade-off between different objective functions, where the light curve presents the largest constraining power and has the smallest value and range among the four $\chi^2$. 

The remaining four artificial binary systems (see columns 2 to 5 in Fig.~\ref{Fig:synthsres}) feature larger relative uncertainties that, in some cases, reach up to 10\% (though deviations from the true parameters are smaller). Furthermore, the final front of solutions shows multiple local minima and a larger scatter. These cases also have a radius-inclination degeneracy, also indirectly reflected in the inferred effective temperature values impacted by uncertainties in the \logg\ parameter through the stellar radius. Yet, the deepest and most narrow minima occur at the right locations, i.e. correspond to the true parameter values. Moreover, for the majority of our synthetic systems, identification of (incorrect) local minima and their rejection proves to be a straightforward task. A notable example is the B2+B2 system, where two distinct minima are present that clearly belong to states where two nearly twin components have been swapped; namely, solutions with the opposite $\omega$ values and swapped $f_c$ and $f_s$ parameters (see \href{https://zenodo.org/records/15388159}{Fig. E.1} for the corner-plots and second column in Fig.~\ref{Fig:synthsres} for the final solution). We also note that while residuals deviate from pure white noise in the light curve solutions in Fig. ~\ref{Fig:synthsmulti_merged}, they do not exceed the 1\% level in their maximum in any of the five simulated cases (and are within 0.5\% in four out of the five system realisations). 

Even though observations for the four out of five artificial systems were faked such as to reflect typical challenges in the data collection and processing, the obtained results are likely the most optimistic scenarios of any real object. In particular, uniform coverage of the orbital cycle with spectroscopic observations is not guaranteed, especially when large samples of stars are observed within a single observational program. Furthermore, both photometric and spectroscopic observations will contain sources of uncertainties other than the well-behaved noise that we assumed in our simulations. For example, space-based photometric observations will show sector-to-sector (in the case of TESS) or quarter-to-quarter ($Kepler$ and PLATO) variations that are subject to imperfect de-trending. On the side of spectra, normalisation to the local continuum is a highly subjective procedure that will deliver data of variable quality depending on factors like personal experience, S/N of the raw spectra, shape of the instrument response function, etc. We expect the above-mentioned factors to lead to larger uncertainties in the location of the correct global minimum, especially when different objective functions (observables) have different relative depths for different minima.

Corner plots provided in \href{https://zenodo.org/records/15388159}{Appendix E (on Zenodo)} and solutions illustrated in Fig.~\ref{Fig:synthsres} represent a snapshot at the final stages of the evolution and a final solution, respectively. A more informative way of looking at the results is to trace the evolution of the entire population for each element in the vector of free parameters. Figure~\ref{Fig:B5_evoM1} shows a temporal evolution (represented by the generation number) of the primary's mass distribution across the population for the 'perfect' case of the B5+B7 artificial system (see Table~\ref{tab:synthetic_systems} for details). The upper panels present all existing values for every generation, showing general parameter space exploration. Lower panels feature several slices at a single snapshot in evolution (generations 0, 10, 50, 100, and 300), showing the histograms of the primary's mass distribution across the population. The same is shown for the primary's \teff\ in Fig. \ref{Fig:B5_evoTeff}. We can see that the algorithm converges rapidly from a uniform distribution in a wide parameter range with an initial Sobol grid towards a single true solution. The convergence is achieved within the first few tens generations and stays within a deep sharp minimum for the remaining set of almost a thousand generations. Yet, the algorithm maintains its exploratory capabilities during these 'stable' generations, which is recognised as 'noise' in local minima in the upper panel. Though some of the local minima are long-lived, the slice histograms barely contain any traces of these minima as most individuals in the population stay in the deepest minimum. The same evolution plots (without detailed slices) are shown for each optimised parameter in \href{https://zenodo.org/records/15388159}{Appendix F}. We note that, in this particular example, the number of individuals in the population equals to 480 for seven dimensions (optimised parameters) with four objective functions. This example shows the efficiency of MOO with the genetic algorithm, which reveals its potential for automated optimisation on multiple objects with computationally expensive on-the-fly generation of synthetic spectra. On the other hand, genetic algorithms tend to converge slowly to the exact minimum location once its general position is found. Therefore, performing a local gradient optimisation might be required to refine the final solution.

\section{Tests on a subsample of well-known observed systems}\label{sec:tests_obs}

In this section, we describe how we tested the implementation of the framework in a setting of realistic observational data. We aimed to achieve a setup that is as close as possible to the intended use on our {\sc hermes} massive EB campaign, described in \citet{tkachenko_observational_2024}. This means combining TESS light curves and eight to ten high-resolution spectra with a (quasi-)uniform coverage of the orbital cycle. This time, we have required a confident literature solution for the validation of our framework. Another requirement for the selection of suitable systems is a diversity in stellar mass. With this in mind, we selected three systems (WW Aur, U Oph, and V453 Cyg) that include the time series of {\sc hermes} spectra and that were previously thoroughly investigated elsewhere  based on both photometric and spectroscopic data. While the WW Aur and V453 Cyg systems have TESS light curves available, the U Oph system was not observed by TESS. Therefore, for the latter system, we use light curves obtained by \citet{vaz_absolute_2007} with the ESO SAT telescope in La Silla, Chile. The same archival light curves were used in the study by \citet{johnston_modelling_2019}, together with time series of {\sc hermes} spectra. Hence, our analysis of the U Oph system relies on exactly the same observational data set as was exploited in \citet{johnston_modelling_2019}. TESS light curves for WW Aur and V453 Cyg were retrieved with the {\sc Lightkurve} package \citep{lightkurve}. For the list of epochs of {\sc hermes} spectra observations with S/N values and phase coverage, see Tables \ref{tab:WWAur_phases},  \ref{tab:UOph_phases},  and \ref{tab:V453Cyg_phases} for the three observed systems.

\subsection{WW Aur  (2 $M_{\odot}$)}\label{sec:wwaur}

WW Aurigae (WW Aur) is a well-studied detached eclipsing binary system consisting of two metallic-lined A-type stars that reside in an orbit of about 2.5 days. High-precision measurements place the mass and radius of the primary (secondary) at about 1.96 (1.81)~$M_{\odot}$ and 1.93 (1.84)~$R_{\odot}$, respectively  \citep{southworth_absolute_2005}. The system exhibits a nearly circular orbit, despite being believed of a fairly young age of about 90 million years \citep{southworth_absolute_2005}. 
Detailed abundance analyses revealed underabundances of CNO elements and significant over-abundances of heavier metals, confirming the Am nature of the stars. \citet{takeda_compositional_2019} reported moderate over-abundances, while \citet{catanzaro_tess_2024} found significantly larger enhancements, consistent with strong metallic peculiarities. This aligns with the earlier conclusion by \citet{southworth_absolute_2005} that stellar models with  high metallicity (Z=0.06) are required to reach an agreement between evolutionary tracks and the system's observed parameters.

\begin{figure}[h!]
            {\includegraphics[clip,width=9.0cm,trim={0.2cm 0.2cm 0.0cm 0.0cm}]{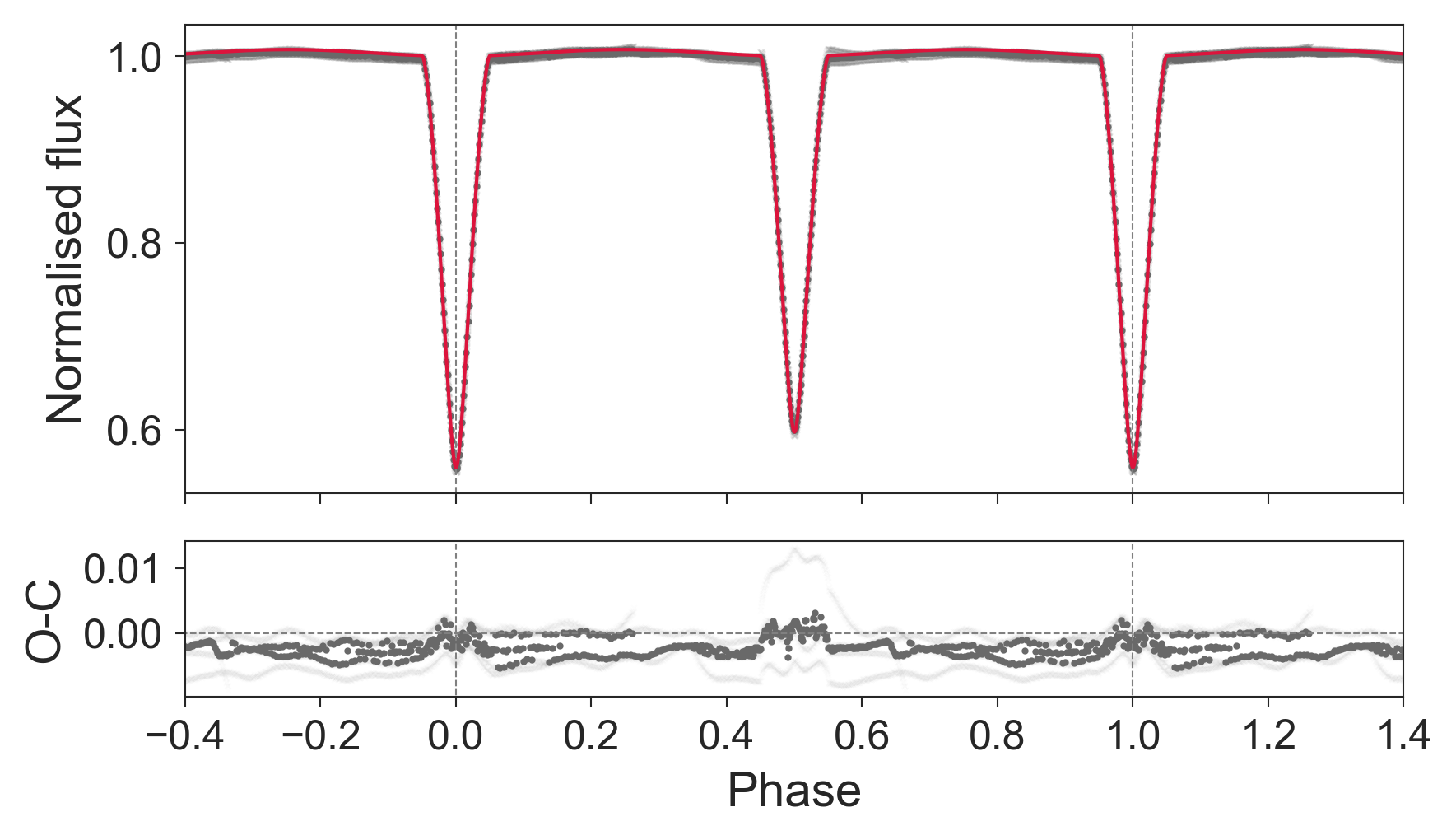} }
      \caption{ Comparison between the best-fit model (red line) and TESS observed TESS (grey dots) light curves of WW Aur. Dark grey dots show the re-binned light curve that was fitted and light grey dots show the original phase-folded light curve.   }
         \label{Fig:WWAur_obsLC}
\end{figure}

\begin{figure*}[h!]
   \centering
\includegraphics[clip,width=500pt,trim={0.2cm 0.9cm 0.0cm 0.2cm}]{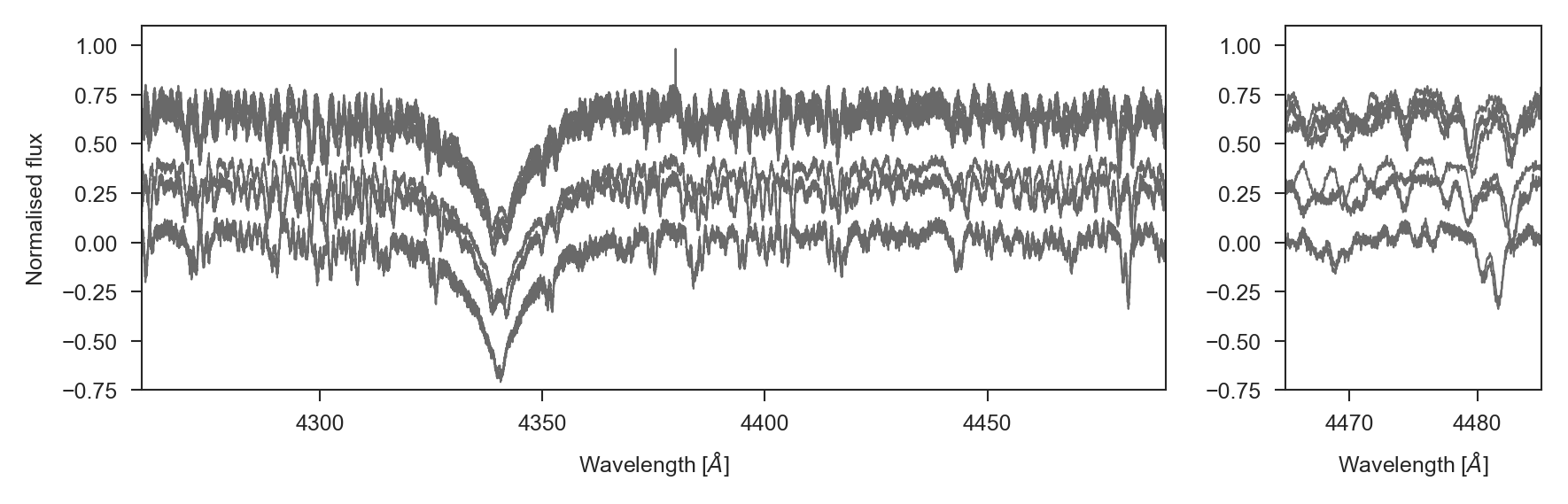}  
\includegraphics[clip,width=500pt,trim={0.0cm 0cm 0.2cm 0.2cm}]{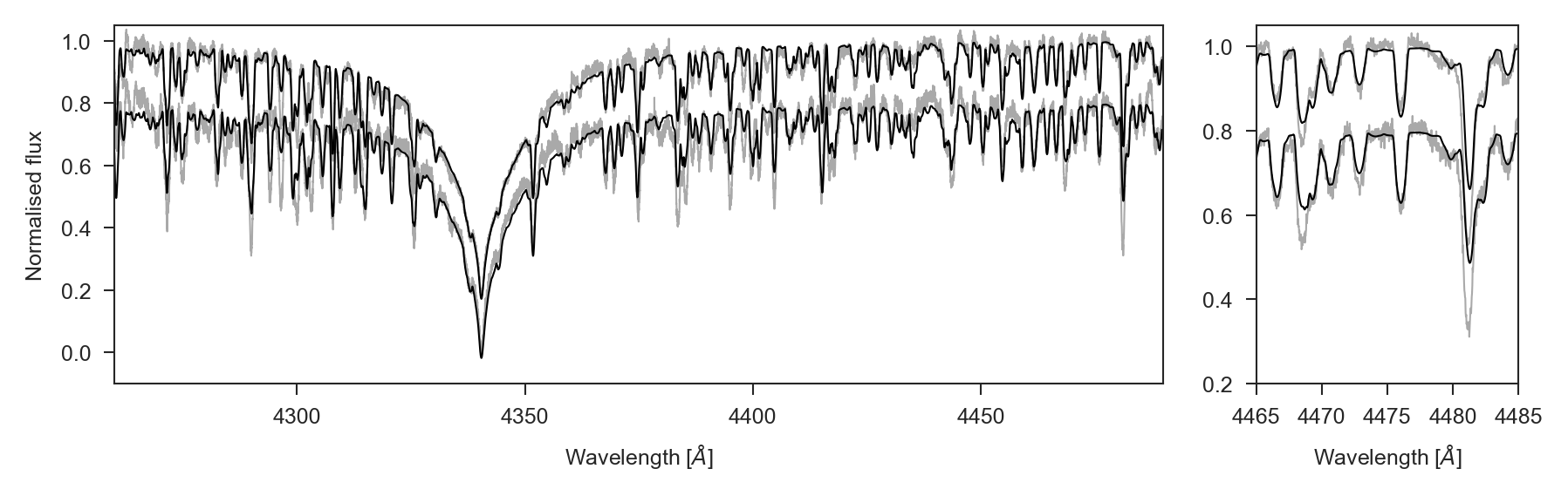}  
      \caption{Top row: Time series of the observed spectra shifted vertically according to their orbital phase value. Right panel shows a zoom-in into the pair of \ion{He}{i}~4471~\AA\ and \ion{Mg}{ii}~4481~\AA\ lines that are also visible at the edge of a wider wavelength range presented in the left panel. Bottom row: Disentangled spectra of both binary components (grey line) together with the best-fit model spectrum (black line).}
    \label{Fig:WWAur_obsSpec}
\end{figure*}

\begin{table}
\centering
\caption{Phase coverage and S/N values of spectroscopic observations of WW Aur (sorted by orbital phase).}
\begin{tabular}{lll}

Phase & BJD & SNR$_{5000\AA}$  \\
\hline

0.262 &  2459900.783 &  96 \\
0.307 &  2459696.371 &  82 \\
0.325 &  2459875.692 &  107 \\
0.354 &  2459691.439 &  75 \\
0.617 &  2459901.680 &  104 \\
0.701 &  2459697.365 &  83 \\
0.711 &  2459876.668 &  118 \\
0.940 &  2459874.720 &  114 \\
0.949 &  2459690.416 &  78 \\

\end{tabular}
\label{tab:WWAur_phases}
\end{table}

Table~\ref{tab:WWAur} features the inferred fundamental absolute parameters of the systems, as derived in \citet[][second column]{southworth_absolute_2005}, \citet[][third column]{wilson_distances_2009}, \citet[][fourth column]{catanzaro_tess_2024} and in this work (column 5). 
In Fig.~\ref{Fig:WWAur_obsLC}, we show the comparison between the best-fit model and the light curve and Fig.~\ref{Fig:WWAur_obsSpec} gives the comparison between the same best-fit model and the spectral time series. 

\begin{figure*}
   \sidecaption
            {\includegraphics[clip,width=12cm,trim={0.0cm 0.0cm 0.0cm 0.0cm}]{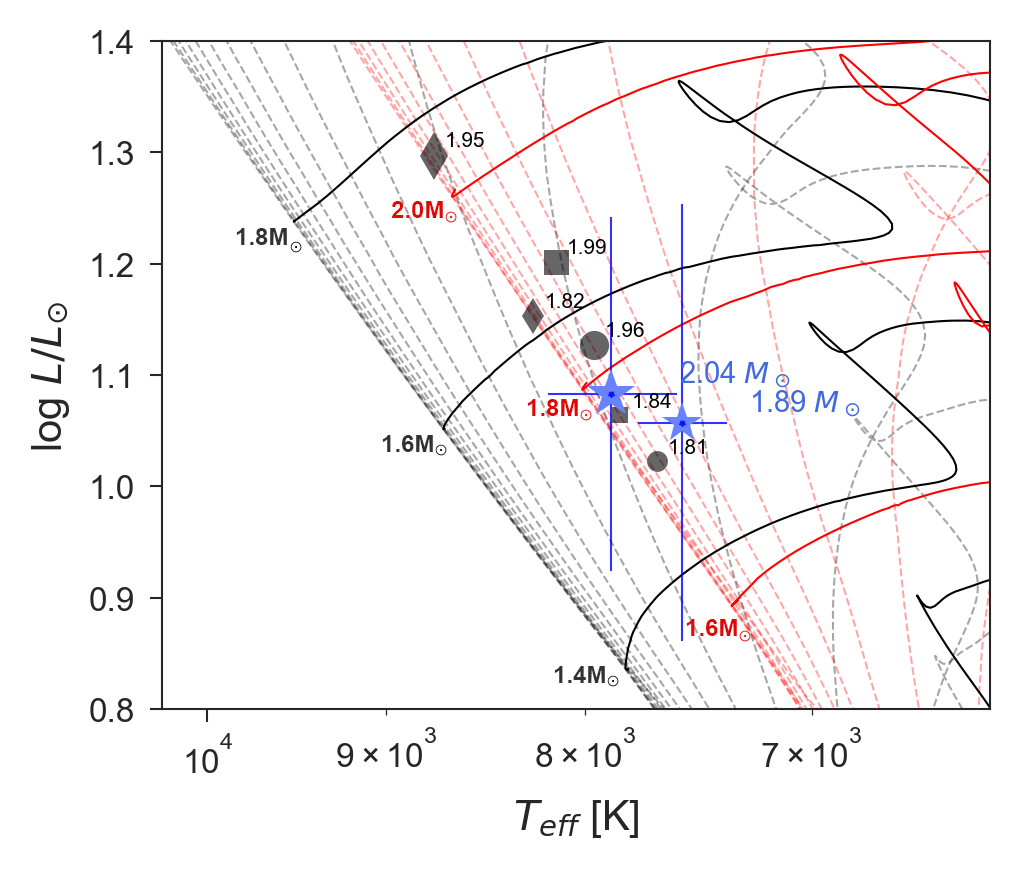}}
      \caption{HR diagram showing the position of WW Aur components. Literature solutions are shown with grey symbols, where circles, squares, and diamonds correspond to columns in Table~\ref{tab:WWAur} respectively. The solution obtained in this study is shown with blue symbols and error bars. Black lines show solar-scaled MIST evolutionary tracks and isochrones for $Z=0.014$, while red lines show tracks and isochrones for enhanced metallicity $Z=0.06$. The dynamical masses corresponding to each point are indicated as text labels. We emphasise that the evolutionary tracks are plotted here for illustrative purposes only and they were not used in the parameter derivation within our framework.  \\ \\ \\ \\  \\ }

         \label{Fig:WWAur_HRD}
\end{figure*}

The HR diagram positioning of the system is shown in Figure~\ref{Fig:WWAur_HRD}, displaying both solar metallicity ($Z=0.014$) and enhanced metallicity ($Z=0.06$) MIST tracks and isochrones. It is evident that all solutions from the literature, as well as our solution, require tracks with a strongly enriched composition to avoid generating a mass discrepancy. However, for Am stars, the surface chemical peculiarities are not expected to reflect the bulk composition, implying that helium abundance should remain close to the solar value even if metals are enhanced at the surface, and the usage of bulk-enriched evolutionary models is not justified. This is what makes WW Aur an interesting but challenging study case. However, the purpose of our study is to test the new methodology, so we further focus on discussing deviations from previously reported parameters and restrain from conclusions considering the evolutionary status of WW Aur. Moreover, we emphasise that evolutionary tracks are only plotted for illustrative purposes; they are not used during the parameter optimisation within our framework in any way and are only needed if we focus on evolutionary masses and ages.

Another noticeable feature in Fig.~\ref{Fig:WWAur_HRD} is that our analysis results in an unequal age solution for the two binary components, unless pre-MS isochrones are considered. We have a strong reason to believe that it is the effective temperature of the secondary that is largely responsible for the unequal age scenario for the system. Indeed, when carefully inspecting the quality of the fit of the H$_{\gamma}$ profile in Fig.~\ref{Fig:WWAur_obsSpec}, its red wing is fitted purely owing to the asymmetry present in the disentangled spectrum. Such artefacts are not unusual in the spectral disentangling, especially in the regions of strong and broad spectral lines such as those of hydrogen in the spectra of main-sequence AF-stars. Attempts to correct for such line asymmetries during the process of spectrum re-normalisation will inflict additional uncertainty in the determination of atmospheric parameters, in particular the effective temperature. The analysis of spectra of WW Aur is further complicated by the fact that both components of the system are known peculiar metallic-lined (Am) stars.

\begin{table*}
\centering
\caption{Properties of WW Aur.}
\begin{tabular}{lcccc}

Property & \citet{southworth_absolute_2005} & \citet{wilson_distances_2009}  & \citet{catanzaro_tess_2024} & This work  \\
\hline

M$_1$ ($M_\odot$)  & $1.964 \pm 0.007$ & $1.989 \pm 0.009$  & $1.95 \pm 0.01$  & $2.035 \pm 0.186$  \\

M$_2$ ($M_\odot$) & $1.814 \pm 0.007$ & $1.838 \pm 0.008$  & $1.82 \pm 0.01$ & $1.889 \pm 0.181$ \\

R$_1$ ($R_\odot$) & $1.927 \pm 0.011$ & $2.008 \pm 0.008$ & $1.94 \pm 0.02$ & $1.872 \pm 0.156$ \\

R$_2$ ($R_\odot$) & $1.841 \pm 0.011$ & $1.850 \pm 0.009$  & $1.85 \pm 0.02$ & $1.973 \pm 0.217$ \\

$T_\text{eff1}$ (K) & $7960 \pm 420$ & $8140 \pm 23$ & $8750 \pm 200$ & $7890 \pm 300$   \\

$T_\text{eff2}$ (K) & $7670 \pm 410$ & $7836 \pm 20$  & $8250 \pm 200$ & $7560 \pm 200$ \\

\end{tabular}
\label{tab:WWAur}
\end{table*}

Even with a poor spectroscopic coverage of the orbital cycle (see Table \ref{tab:WWAur_phases} or gaps in vertical shifts of spectra in Fig. \ref{Fig:WWAur_obsSpec}), the solution obtained by us does not differ from those reported in \citet{southworth_absolute_2005} and \citet{wilson_distances_2009} by more than they differ from each other. We also note that the differences between parameter values obtained by us and those reported in the literature are well within 1$\sigma$ uncertainties quoted by us in Table~\ref{tab:WWAur}. The exception are temperatures in \citet{catanzaro_tess_2024}, which are significantly higher than all other reported solutions, which also lead to a higher position of the said solution on HRD in \ref{Fig:WWAur_HRD}. Higher temperatures in \citet{catanzaro_tess_2024} may be attributed to differences in treatment of metallicity; specifically, the authors assumed Solar abundances and metallicity to compute atmosphere models, and used solar metallicity tracks to compare binary parameters to evolutionary properties. On the contrary, we fixed the metallicity at a higher value of [Fe/H]=0.5 in spectra synthesis, and plot solutions on top of evolutionary tracks with the same [Fe/H]=0.5, following conclusions of \citet{southworth_absolute_2005}. Note that known peculiarities in abundances make both approaches not ideal as varying individual abundances is required rather than scaling Solar abundances. The Am nature of the binary components biases the determination of their effective temperatures and leads to significant uncertainties, even when the rest parameters are strongly consistent between different studies. 

The reported uncertainties are substantially larger in our study than uncertainties reported in any of the above-mentioned studies, however. Such an increase in uncertainties is associated with the inclusion of all data and parameters simultaneously, allowing for a greater freedom for parameters to compensate for each other. This is in contrast to fixing some of the dimensions in the traditional approach. In other words, our method gives estimation of systematic uncertainty by demanding a consistent solution, while such systematics are pretended to be absent from an iterative approach that treats the light curve and spectra separately.

Due to the nature of MOO (see Section~\ref{sec:multi-objective} for discussion), it provides a cloud of solutions, laying on an optimal trade-off surface between different competing objective functions (in our case: light-curve fitting, spectral disentangling, and fitting the two disentangled spectra). In this case, the definition of the single best solution lay on the user: we may define weights between objectives based on each data quality and completeness, may follow isochrones of the models, or look through all solutions and select a compromise. WW Aur serves an excellent example: although there are solutions that align perfectly with the MIST isochrones, we discarded them based on the dissatisfying quality of the light curve fit and/or the presence of the mass discrepancy. Instead, we preferred a solution that satisfied the following criteria: a better agreement between the dynamical and evolutionary masses, and high quality of spectral disentangling with flat residuals. The MOO subjects the user to a need for making an informed choice, where we can evaluate all good solutions (equally good for the algorithm that is not aware of nuances of different datasets or physics) and decide which dataset is more reliable and which fits deficiencies may be tolerated more than others. For example, instrumental artefacts and poorly fitted lines in the spectra are treated by the algorithm as equally important deviations from models, while they can be assessed differently by the user. While it seems like a complication compared to the traditional approach, the latter is not free of that choice either, but it is hidden behind selecting initial assumptions and constraints, as well as the scheme of iterative treatment of different data types. At the same time, a unified self-consistent framework provides an efficient way to explore the parameter space and is more robust in terms of quantifying underlying uncertainties.

\subsection{U Oph  (5 $M_{\odot}$)}\label{sec:uoph}

U Ophiuchi (U Oph) is another well-studied detached eclipsing binary system comprising two main-sequence stars of spectral type B5V. The system has been investigated extensively due to its proximity ($\sim$ 186 pc) and its importance as a benchmark for stellar evolution models of intermediate-mass stars. Mass and radius estimates of the primary and secondary components show some variation across studies, with primary masses ranging from 4.93 to 5.27 $M_{\odot}$ and secondary masses from 4.56 to 4.78 $M_{\odot}$. Radii estimates vary from 3.29 to 3.48 $R_{\odot}$  for the primary and 3.01 to 3.11 $R_{\odot}$ for the secondary \citep{holmgren_absolute_1991, vaz_absolute_2007, budding_absolute_2009, johnston_modelling_2019}. The system has a nearly circular orbit and exhibits apsidal motion with a period around 21 years. This apsidal motion is attributed to the presence of a distant third companion with a mass around 1 $M_{\odot}$. In our analysis, we did not account for apsidal motion as it is negligible on the time span of observations.

\begin{table}
\centering
\caption{Phase coverage and S/N values of spectroscopic observations of U Oph (sorted by orbital phase).}
\begin{tabular}{lll}

Phase & BJD & SNR$_{5000\AA}$  \\
\hline

0.066 & 2457610.503 & 114  \\
0.089 & 2457553.512 & 138  \\
0.196 & 2457501.693 & 160  \\
0.294 & 2457508.568 & 151  \\
0.421 & 2457614.454 & 148  \\
0.497 & 2457505.553 & 155  \\
0.607 & 2457611.410 & 151  \\
0.609 & 2457564.448 & 115  \\
0.694 & 2457507.560 & 160  \\
0.796 & 2457502.699 & 162  \\
0.894 & 2457509.574 & 154  \\
0.997 & 2457494.650 & 151  \\

\end{tabular}
\label{tab:UOph_phases}
\end{table}

\begin{figure}
            {\includegraphics[clip,width=9.2cm,trim={0.2cm 0.2cm 0.0cm 0.0cm}]{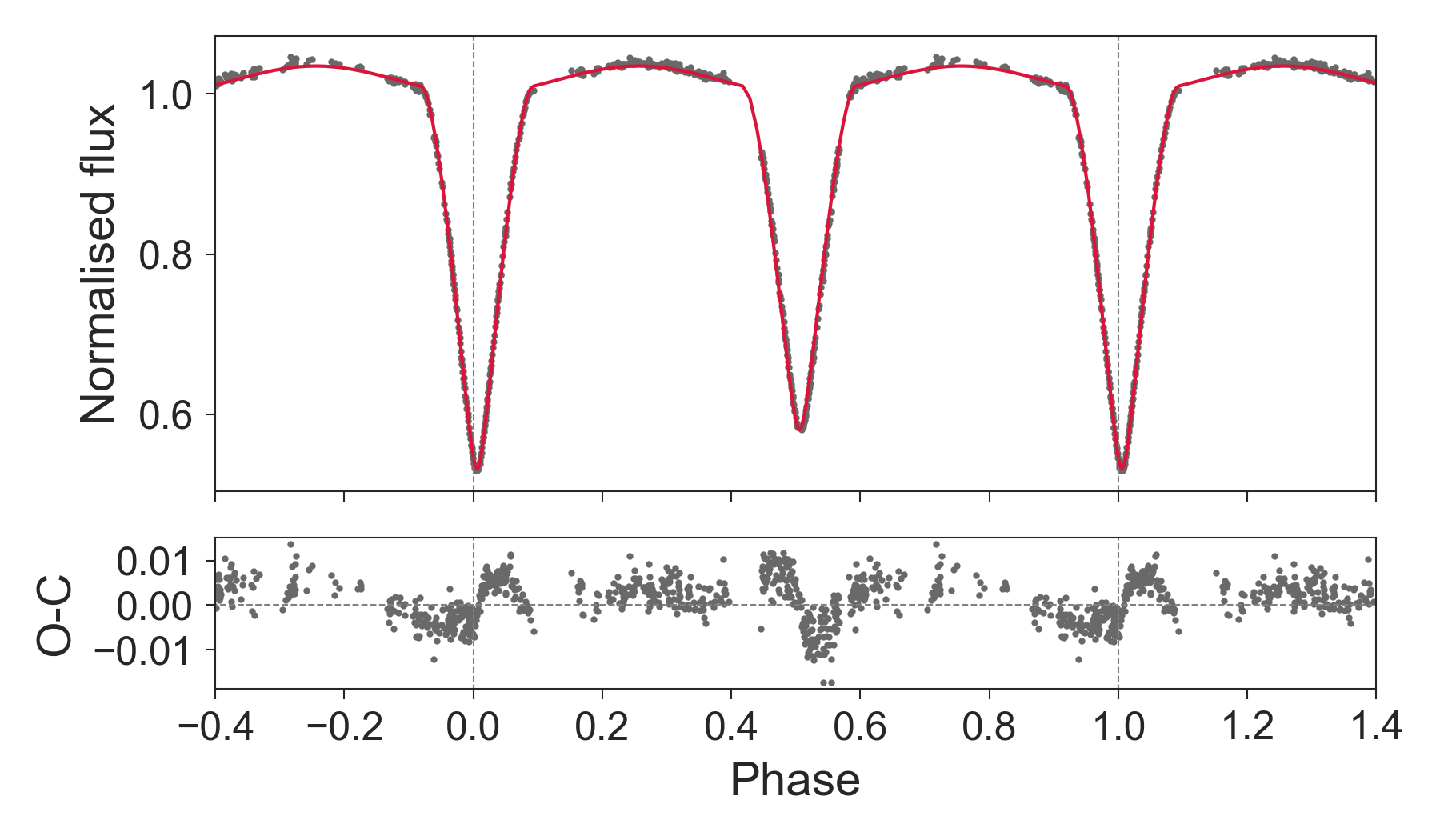} }
      \caption{ Same as in Figure~\ref{Fig:WWAur_obsLC}, but for the U Oph system.}   
         \label{Fig:UOph_obsLC}
\end{figure}

\begin{figure}
            {\includegraphics[clip,width=9.5cm,trim={0.4cm 0.0cm 0.0cm 0.0cm}]{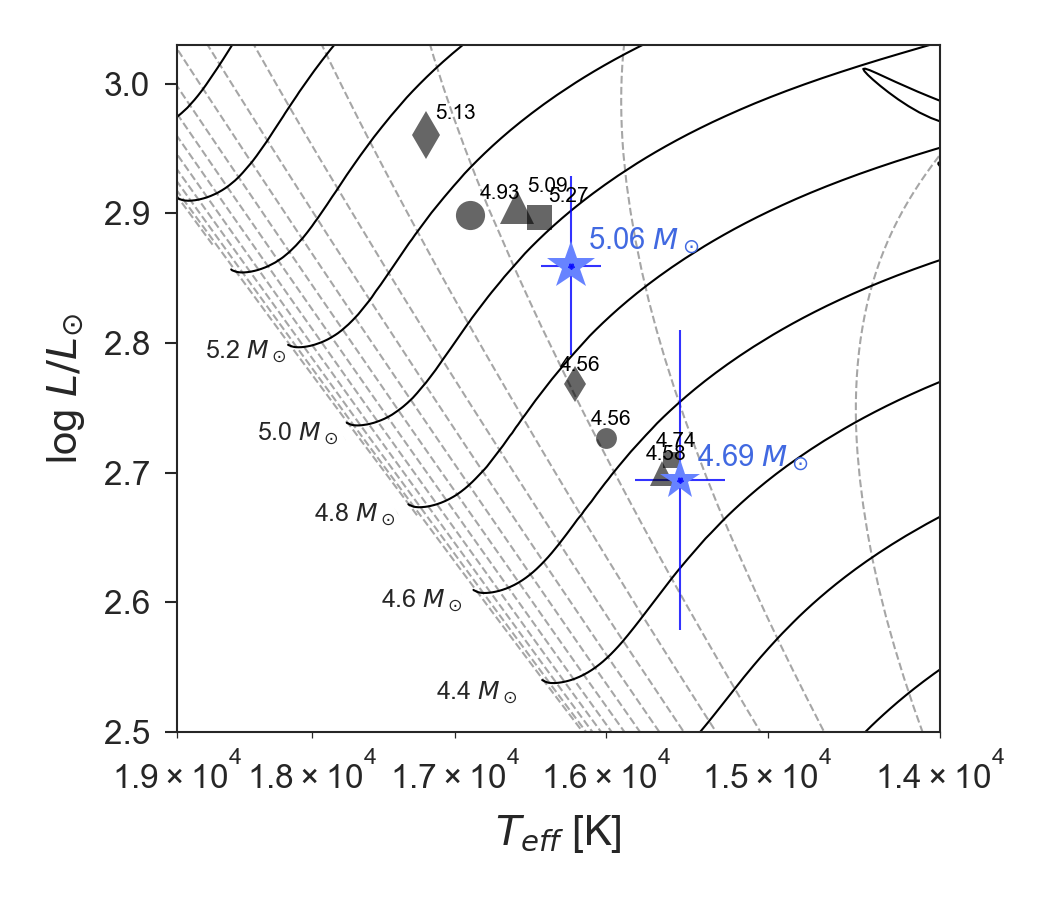}}
      \caption{HR diagram showing the position of U Oph components. Literature solutions are shown with grey symbols, where circles, squares, diamonds, and triangles correspond to columns in Table~\ref{tab:UOph} respectively. The solution obtained in this study is shown with blue symbols and error bars. Black lines show MIST evolutionary tracks and isochrones for $Z=0.014$. Solar composition was also adopted during spectral synthesis. The dynamical masses corresponding to each point are indicated as text labels. }  

         \label{Fig:UOph_HRD}
\end{figure}

A comparison of the literature solutions with the solution obtained in this study is presented in Table \ref{tab:UOph}; a graphical representation is provided in Figs. \ref{Fig:UOph_obsLC}, \ref{Fig:UOph_HRD}, and \ref{Fig:UOph_obsSpec}. With a decent, close to uniform spectroscopic coverage of the orbital cycle, U Oph shows an excellent agreement with the literature solutions as well as with MIST the tracks and isochrones (in contrast to WW Aur).

\begin{table*}
\caption{Properties of U Oph.}
\centering
\begin{tabular}{lccccc}
Property & \citet{holmgren_absolute_1991} & \citet{vaz_absolute_2007} & \citet{budding_absolute_2009} & \citet{johnston_modelling_2019} & \textbf{This work} \\
\hline
M$_1$ ($M_\odot$) & $4.93 \pm 0.05$ & $5.27 \pm 0.09$ & $5.13 \pm 0.08$ & $5.09 \pm 0.06$ & \textbf{$5.06 \pm 0.24$} \\
M$_2$ ($M_\odot$) & $4.56 \pm 0.04$ & $4.74 \pm 0.07$ & $4.56 \pm 0.07$  & $4.58 \pm 0.05$ & \textbf{$4.67 \pm 0.22$} \\
R$_1$ ($R_\odot$) & $3.29 \pm 0.06$ & $3.47 \pm 0.02$ & $3.41 \pm 0.03$ & $3.44 \pm 0.01$ & \textbf{$3.41 \pm 0.26$} \\
R$_2$ ($R_\odot$) & $3.01 \pm 0.05$ & $3.11 \pm 0.03$ & $3.08 \pm 0.03$ & $3.05 \pm 0.01$ & \textbf{$3.07 \pm 0.30$} \\
$T_\text{eff1}$ (K) & $16900 \pm 1500$ & $16600 \pm 400$ & 17200 & $16580 \pm 180$ & \textbf{$16230 \pm 200$} \\
$T_\text{eff2}$ (K) & $16000 \pm 1500$ & $15500 \pm 400$ & 16200 & $15650 \pm 200$ & \textbf{$15540 \pm 280$} \\

\end{tabular}
\label{tab:UOph}
\end{table*}

\begin{figure*}
   \centering
\includegraphics[clip,width=180mm,trim={3cm 0cm 2cm 0cm}]{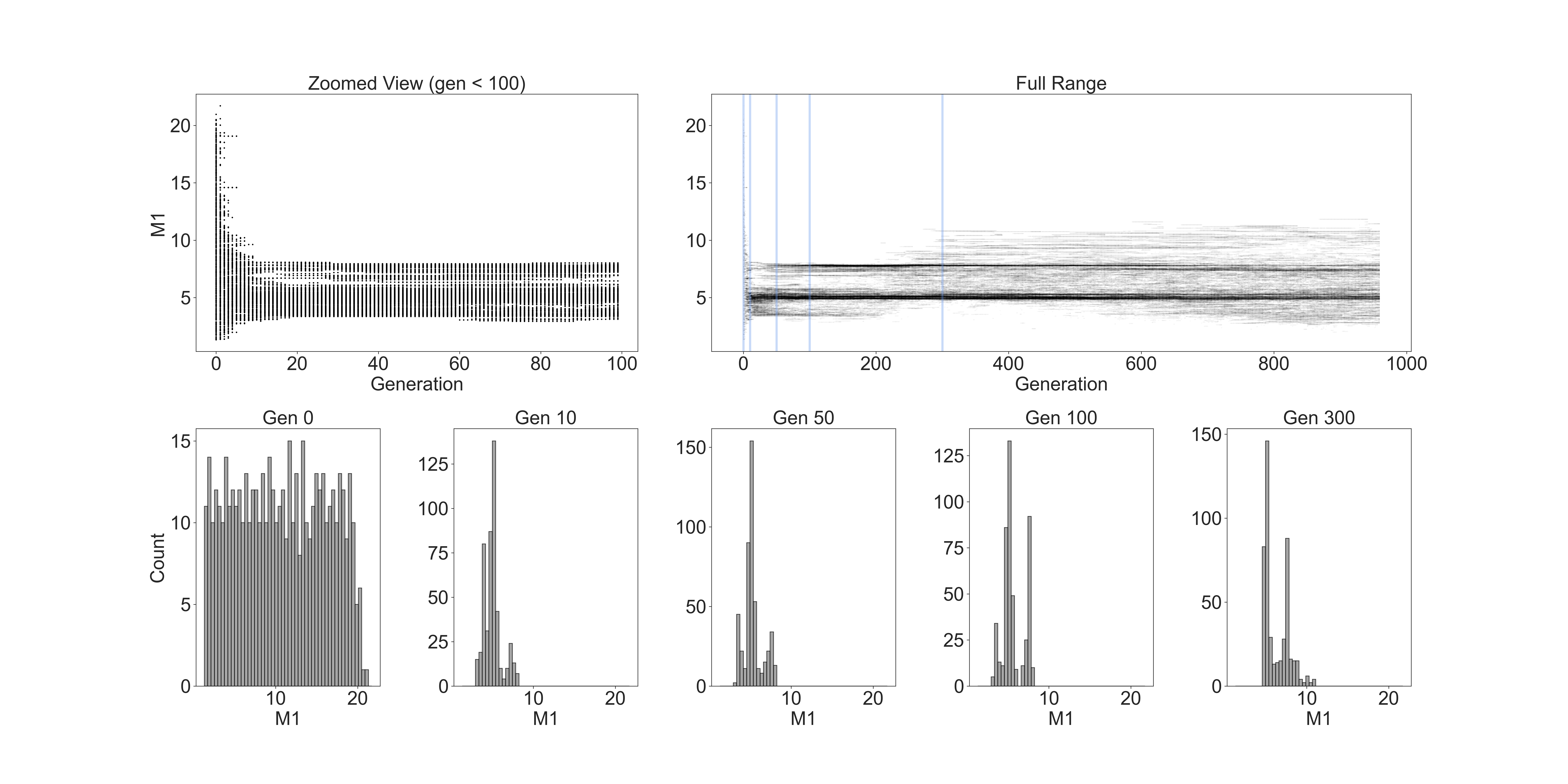}  
      \caption{ Same as Fig. \ref{Fig:B5_evoM1}, but for the case of real observed data of U Oph.  }
    \label{Fig:UOph_evoM1}
\end{figure*}

\begin{figure*}
   \centering
\includegraphics[clip,width=180mm,trim={3cm 0cm 2cm 0cm}]{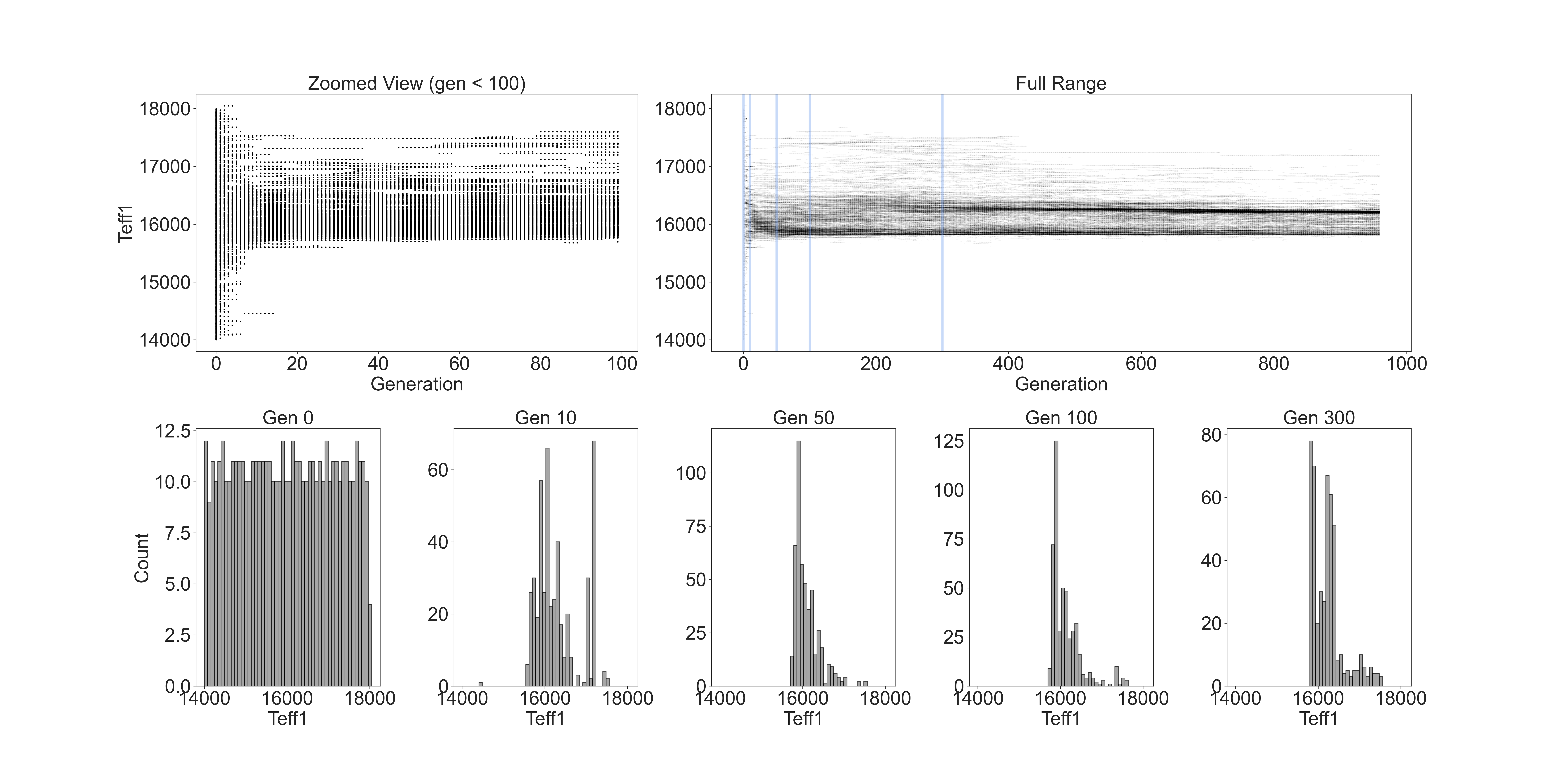}  
      \caption{ Same as Fig. \ref{Fig:UOph_evoM1}, but for \teff.  }
    \label{Fig:UOph_evoTeff}
\end{figure*}

By analogy with the tests on artificial binary systems, Figs.~\ref{Fig:UOph_evoM1} and \ref{Fig:UOph_evoTeff} present the evolution of the primary's mass and effective temperature distributions along the population. \href{https://zenodo.org/records/15388159}{Fig. F.3} provides a summary of the evolution of the optimised parameters for the U Oph system. Compared to the case of the B5+B7 artificial binary system (see Table~\ref{tab:synthetic_systems} for parameters of the system and Figs. \ref{Fig:B5_evoM1}, \ref{Fig:B5_evoTeff}, and \href{https://zenodo.org/records/15388159}{F.1} for the evolution of the optimised parameters), the landscape of the objective functions is more complex for the U Oph system. In particular, the population is spread out within several local minima of comparable depths, and the minima are not as narrow as they were in the case of the artificial binary. The highest density corresponds to a solution that provides the best trade-off found across all four objectives, i.e. light curve, spectral disentangling, and fitting of the components' disentangled spectra. All other solutions may only improve some of the objectives at the cost of the others. Selecting a solution based on the highest density cluster location seems a natural choice, provided we equally trust the models and data used in each of the four objectives.

\subsection{V453 Cyg  (14 $M_{\odot}$)}\label{sec:v453cyg}

V453 Cygni (V453 Cyg) is a high-mass eclipsing binary system with components of B0 spectral type. The system components' masses and radii are well-established, with values around 13.9 $M_\odot$  and 8.6 $R_\odot$ for the primary, and 11.1 $M_\odot$ and 5.5 $R_\odot$ for the secondary \citep{pavlovski_physical_2018}.
V453 Cyg is known for the pronounced mass discrepancy problem where the dynamical and evolutionary masses are found to differ by 30\%. Additionally, recent TESS observations revealed $\beta$ Cep-type pulsations in the primary component of the system \citep{southworth_discovery_2020}. 

\begin{table}
\centering
\caption{Phase coverage and S/N values for spectroscopic observations of V453 Cyg (sorted by orbital phase).}
\begin{tabular}{lll}

Phase & BJD & SNR$_{5000\AA}$  \\
\hline

0.164 & 2459417.445 & 99 \\
0.178 & 2459421.391 & 104 \\
0.220 & 2459417.665 & 113 \\
0.222 & 2459425.454 & 102 \\
0.243 & 2459421.644 & 87 \\
0.258 & 2459425.591 & 102 \\
0.419 & 2459418.436 & 101 \\
0.453 & 2459422.460 & 95 \\
0.474 & 2459426.431 & 98 \\
0.478 & 2459418.668 & 101 \\
0.502 & 2459430.432 & 116 \\
0.506 & 2459422.668 & 103 \\
0.530 & 2459426.648 & 98 \\
0.531 & 2459430.543 & 115 \\
0.569 & 2459430.692 & 101 \\
0.675 & 2459419.434 & 70 \\
0.696 & 2459419.515 & 96 \\
0.705 & 2459423.442 & 105 \\
0.719 & 2459427.386 & 99 \\
0.729 & 2459419.644 & 72 \\
0.747 & 2459431.383 & 110 \\
0.755 & 2459427.525 & 105 \\
0.763 & 2459423.666 & 103 \\
0.786 & 2459427.646 & 103 \\
0.935 & 2459420.446 & 101 \\
0.949 & 2459424.391 & 108 \\
0.976 & 2459428.384 & 95 \\
0.990 & 2459420.659 & 97 \\
 
\end{tabular}
\label{tab:V453Cyg_phases}
\end{table}

It was required to include the reflection and gravity darkening effects in our analysis of the light curve of V453 Cyg. This implies an increase in the number of optimised parameters by at least (for a simplified reflection model) four (i.e. two per binary component).  In addition, the system is known to be slightly eccentric (with e=0.02); hence, an additional variation of $f_c$, $f_s$, and $T_0$ would also be required for the purpose of testing the robustness of our framework. Altogether, we need to optimise 14 parameters for the V453 Cyg system: the ones listed above, the remaining orbital parameters, as well as the radii and temperatures of both components. This high dimensionality poses a challenge for the population size limited to several hundreds individuals, hence we double it for the specific case of V453 Cyg. In addition to that, to help proper parameter space exploration, we initiate a second consequent run with a decreased number of parameters, by fixing eccentricity-related parameters, as well as the reflection and gravity darkening coefficients. The respective parameters are assigned values based on the outcome of the previous run with a full set of 14 free parameters.

\begin{figure}
            {\includegraphics[clip,width=9.2cm,trim={0.2cm 0.2cm 0.0cm 0.0cm}]{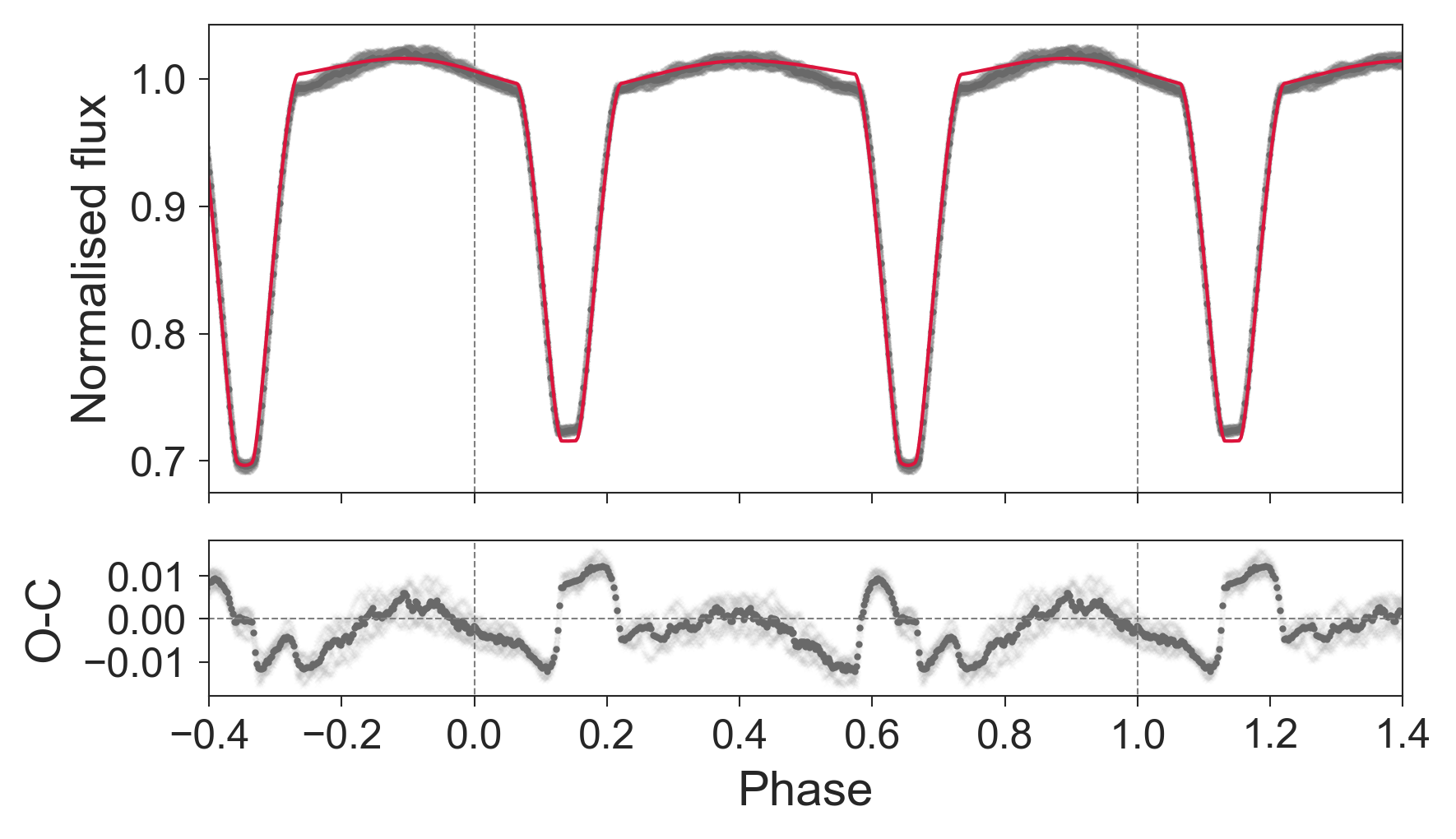} }
      \caption{ Same as in Fig. \ref{Fig:WWAur_obsLC} but for the V453 Cyg system. } 
         \label{Fig:V453Cyg_obsLC}
\end{figure}

Figures \href{https://zenodo.org/records/15388159}{F.4} and \href{https://zenodo.org/records/15388159}{F.5} feature temporal evolution of the primary's mass and effective temperature distributions, respectively. The parameter distributions are presented for both the two above-mentioned runs, i.e. with a full set of 14 free parameters and a limited set of free parameters in the top and bottom panels, respectively. In both cases, the majority of individuals within the population converge to $M_1\approx$14 $M_\odot$ and $T_{\text{eff}1}\approx$ 28000~K. We can see though a wide spread of the population along $T_{\text{eff}1}$, which suggests a complex landscape in the objectives along this parameter and, hence. a large uncertainty.

\begin{table*}
\centering
\caption{Properties of V453 Cyg.}
\begin{tabular}{lcccc}

\hline
Property & \citet{southworth_eclipsing_2004}  & \citet{pavlovski_physical_2018} & \textbf{This work (a)} & \textbf{This work (b)}  \\
\hline

M$_1$ ($M_\odot$) & $14.36 \pm 0.20$ & $13.90 \pm 0.23$ & $ 13.87 \pm 0.24 $ & $ 13.96 \pm 0.24 $ \\

M$_2$ ($M_\odot$) & $11.11 \pm 0.13$ & $11.06 \pm 0.18$ & $ 10.93 \pm 0.29 $ & $ 11.37 \pm 0.29 $ \\

R$_1$ ($R_\odot$) & $8.551 \pm 0.055$ & $8.62 \pm 0.09$ & $ 8.93 \pm 0.27 $ & $ 9.00 \pm 0.27 $ \\

R$_2$ ($R_\odot$) & $5.489 \pm 0.063$ & $5.45 \pm 0.08$ & $ 5.56 \pm 0.11 $ & $ 5.58 \pm 0.11 $  \\

$T_\text{eff1}$ (K) & $26600 \pm 500$ & $28800 \pm 500$ & $ 25990 \pm 520 $ & $ 28060 \pm 520 $  \\

$T_\text{eff2}$ (K) & $25500 \pm 800$ & $27700 \pm 600$ & $ 24500 \pm 730 $ & $ 26710 \pm 730 $  \\

\hline
\end{tabular}
\label{tab:V453Cyg}
\end{table*}

\begin{figure}
            {\includegraphics[clip,width=9.2cm,trim={0.2cm 0.2cm 0.0cm 0.0cm}]{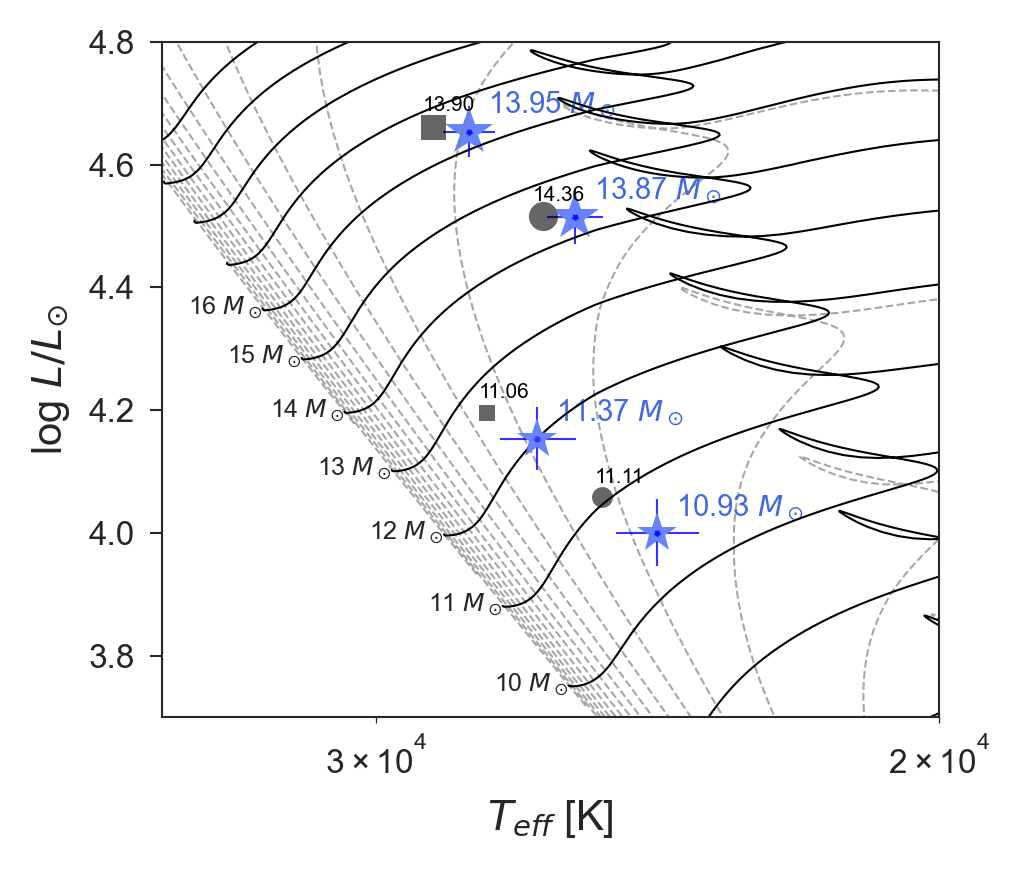}}
      \caption{Same as in Figure~\ref{Fig:UOph_HRD}, but for the V453 Cyg system. Literature solutions are shown with grey symbols, where circles and squares correspond to   \citet{southworth_eclipsing_2004} 
 and \citet{pavlovski_physical_2018} from Table~\ref{tab:V453Cyg}, respectively. Both solutions obtained in this study are shown with blue symbols and error bars (see text for details). } 

         \label{Fig:V453Cyg_HRD}
\end{figure}

\begin{figure*}
   \centering 
\includegraphics[clip,width=500pt,trim={0.1cm 0cm 0.25cm 0.2cm}]{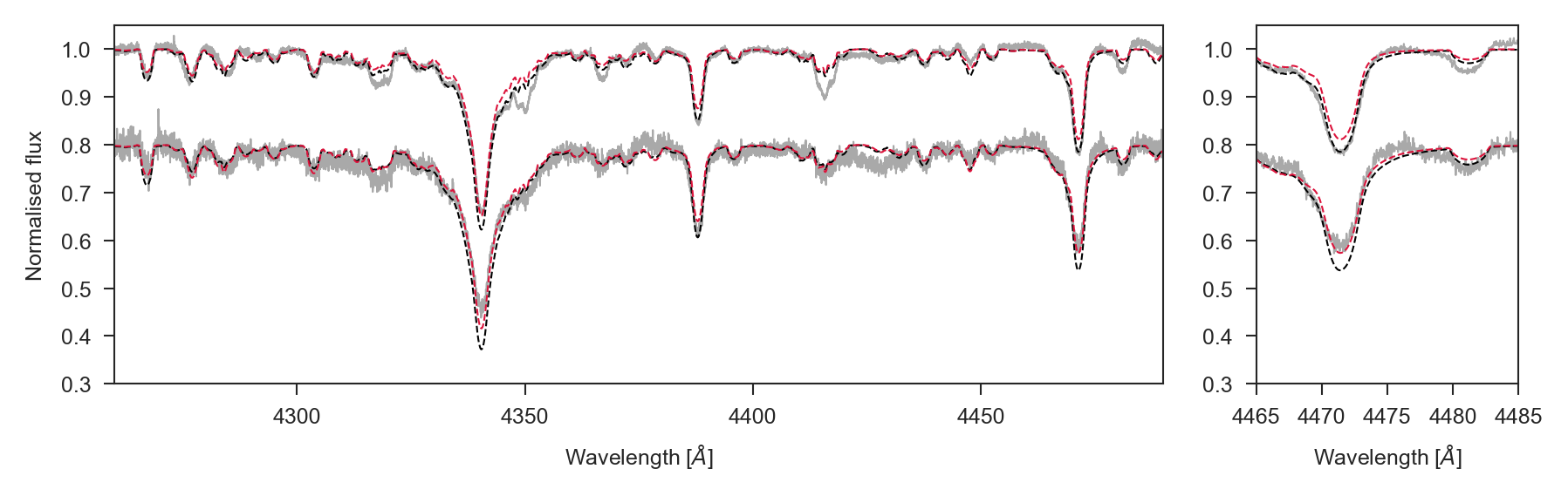}  
      \caption{Comparison between the disentangled spectra of both components of the V453 Cyg system and two different models in each case. Black dashed line features the model spectrum for our solution presented in Table \ref{tab:V453Cyg} and Figs. \ref{Fig:V453Cyg_obsLC}-\ref{Fig:V453Cyg_HRD} and corresponding to lower effective temperatures of 25990 K and 24500 K for the primary and secondary, respectively. Red dashed line shows model spectra corresponding to higher effective temperature values of 28800~K (primary) and 27700~K (secondary), following \citet{pavlovski_physical_2018}. }
    \label{Fig:V453Cyg_comp_spec}
\end{figure*}

What makes the case of the V453 Cyg system particularly interesting in the context of this study is a local minimum that is consistently found in both runs.  The same solution was reported two decades ago by \citet{southworth_eclipsing_2004}. This solution, shown in Fig. \ref{Fig:V453Cyg_obsLC} and Fig. \ref{Fig:V453Cyg_obsSpec}, is characterised by a lower value of the effective temperature of the primary, namely, $T_{\text{eff}1}\approx$26000~K, and a lack of the mass discrepancy for the system. While the dynamical masses of both components remain consistent with the measurements presented in \citet[][see Table~\ref{tab:V453Cyg}]{pavlovski_physical_2018}, the components' positions in the HR-diagram are shifted diagonally towards lower effective temperature and luminosity values (see Fig.~\ref{Fig:V453Cyg_HRD}). At the same time, the surface brightness ratio is preserved. We show both plausible solutions in Fig. \ref{Fig:V453Cyg_HRD} and Table \ref{tab:V453Cyg}. We can see that these two solutions roughly correspond to previously reported solutions.
Although there is a difference of almost 3000~K between the most different solutions (This work (a) and \citet{pavlovski_physical_2018}), which is large even for high-mass stars, it is not necessarily as prominent in the reference frame of spectra. Figure~\ref{Fig:V453Cyg_comp_spec} presents a comparison between synthetic spectra corresponding to solutions of $T_{\text{eff}1}$/$T_{\text{eff}2}$ = 25590/24500~K (black dashed line) and $T_{\text{eff}1}$/$T_{\text{eff}2}$ = 28800/27700~K (red dashed line), and the disentangled spectra of both components (solid lines). The main differences between the two solutions are seen in the depths of hydrogen and helium lines, and it is difficult to judge in some cases which solution is a better one. For example, while \ion{He}{i}~4471~\AA\ line is fitted better at a lower value of $T_{\text{eff}1}$ in the primary's spectrum, the same line is represented better by a higher value of $T_{\text{eff}2}$ in the spectrum of the secondary (right panel in Figure~\ref{Fig:V453Cyg_comp_spec}). We also note that the differences between the two model spectra are of a smaller scale than artefacts in the observed spectra left over from the processes of spectrum normalisation and disentangling. 

The accuracy and precision of the effective temperature inference are further altered by the light dilution factors and departures from LTE that enter the analysis at stages of the spectral disentangling and analysis of the disentangled spectra, respectively. Therefore, deciding between two or more local minima emerging during the optimisation process is a matter of active, informed, and transparent choice. For example, as demonstrated above for the case of V453 Cyg, we can find arguments for the most populated minimum characterised by the best quality of spectral fitting but large mass discrepancy for both binary components. On the other hand, arguments can also be presented for a solution without the mass discrepancy problem but small-scale, local deviations of model spectra from their observed disentangled counterparts. 

To address the challenges of solution selection in multi-objective optimisation, we developed a GUI for post-processing to assist the user in navigating the results of the automated multi-objective search. Not only does the GUI offer a visualisation of the distribution of solutions across both parameter and objective spaces, it also allows for an interactive exploration of any selected solution with on-the-fly computation of the full model with light curve, disentangling and spectral synthesis. Additionally, it (optionally) illustrates metrics of constraints imposed by models of stellar structure and evolution, such as the age difference between binary  components or their mass discrepancies relative to evolutionary models (see the upper panel of Fig.~\ref{Fig:screen1-2}). While these metrics are not included in the optimisation process itself (to avoid setting a dependency on SSE models), they still provide critical context during post-processing and solution selection, enabling  users to make an informed decision. As demonstrated for V453 Cyg, such tools allow us to balance competing metrics and prioritise solutions that align best with a specific understanding of the biases in the observed data of a given object or biases due to any deficiencies in simplified models. Moreover, apart from showing diagnostic plots for the completed optimisation, the post-processing GUI allows for a local single-objective optimisation with the user-defined weights for objectives. This option offers an extra degree of freedom for the user should they want to experiment with the inclusion or elimination of certain free parameters from the fit (see the bottom panel of Figure.~\ref{Fig:screen1-2}).

\section{Conclusions}\label{sec:conclusions}

This study represents a step forward in investigating the mass discrepancy problem in eclipsing binaries by introducing a self-consistent framework that simultaneously models photometric and spectroscopic time series. Through the integration of light curve synthesis, spectral disentangling, and spectral synthesis into a unified model, we approached assessing the role of biases introduced together with the traditional methodology of iterative data refinement. 

The framework named Revising Eclipsing binaries analysis: Photometry And spectroscopy Infused Recipe (\texttt{re:pair}) is designed to incorporate existing techniques (which are, however, often used independently) into a concurrent flow, where a single set of parameters defines all models for different types of observational data. In particular, the framework eliminates the need to iterate between the analyses of independent observational data sets (e.g. light curves versus spectra) and maintains the physical connection between all its free parameters. This becomes possible thanks to the integration of on-the-fly spectral synthesis and a consistent use of the same set of atmosphere models for fitting stellar spectra and light curves, while avoiding intermediate and imprecise steps of using tabulated coefficients.  

Dealing with several types of observational data at the same time required solving an optimisation problem where each point in the parameter space is associated with four objective functions. We applied a multi-objective optimisation to avoid collapsing different objectives into a single value, which would require an assignment of arbitrary weights. Instead, the multi-objective optimisation aims to improve all objectives with no weighting, resulting in a cloud of solutions approaching the optimal trade-off surface. This technique makes it possible to carry out an efficient parameter space exploration in a setup with several conflicting objective functions.

Applications to synthetic test data sets have demonstrated the framework's ability to accurately recover system parameters under realistic observational conditions. Despite challenges such as partial eclipses, low S/N values for some of the spectra in the time series, and heavily blended spectral lines, the \texttt{re:pair} framework provided robust solutions, with uncertainties naturally increasing for more complex systems. These tests also highlighted degeneracies between various parameters, illustrating the importance of exploring parameter space exhaustively to locate global minima while maintaining consistency across datasets.

The analysis of real binary systems offered insights into both the strengths and limitations of the framework. For the cases of WW Aur and U Oph, solutions found in this study were consistent with the literature results. However, both these cases revealed challenges compared to the idealistic synthetic dataset and the traditional iterative approach, such that our underlying physical model appeared too simplistic to provide perfect fits for the observational datasets altogether. At the same time, V453 Cyg, a system with one of the largest mass discrepancies reported in the literature, has showcased the framework’s ability to offer alternative interpretations. In particular, thanks to its ability to efficiently explore objectives space and converge to different solutions simultaneously, the framework recovered an alternative local minimum characterised by lower values of the effective temperature for both binary components. This solution might solve the mass discrepancy in the V453 Cyg system; howevr, it also emphasises the complexity of working with high-mass stars and disentangled spectra in the presence of artefacts in observational data.

Implications for the mass discrepancy problem are twofold. First, our findings hint at part of the mass discrepancy originating from inconsistencies in the traditional data analysis methods, rather than intrinsic deficiencies in stellar evolution models. This highlights the need for standardised, unbiased methodologies, particularly as studies shift towards larger samples. Second, the framework offers a scalable solution for mapping the mass discrepancy across a diverse parameter space, enabling robust comparisons of the dynamical and evolutionary masses without relying on fragmented, object-by-object analyses.

Despite these advances, challenges remain. The framework's computational demands, especially for wide spectral coverage or complex systems, require significant resources and may limit its applicability for large samples. With on-the-fly spectral synthesis, one model computation takes four seconds on a single core (varying with the length of the spectral region), which accumulates to the complete optimisation routine in order of a day(s) per object. 
Additionally, the framework is constrained by the quality of input data. For example, spectral disentangling remains sensitive to the quality of normalisation of the observed spectra. In addition, its robustness depends on how uniform the orbital cycle is covered with the observed time series. Future developments that include (but are not limited to the incorporation of more accurate and physically motivated light curve models) and accounting for spectral variability could further enhance the framework's precision and applicability.

A key challenge in multi-objective optimisation (MOO) is the selection of the most appropriate solution from the Pareto-optimal set, as the algorithm provides a cloud of trade-offs rather than a single 'best' solution. This necessitates an informed choice by the user, balancing competing objectives such as the light curve fitting, spectral disentangling, and spectral fitting for the atmospheric parameter determination. To help address this challenge, we developed a post-processing GUI that enables a detailed exploration of the solution space. The tool allows the user to visualise and evaluate the impact of different trade-offs, making the decision process more transparent and grounded in both data quality and physical context. While this decision-making step might seem like a limitation of the framework, it is equally present in the traditional iterative approaches; in particular, it is merely obscured by assumptions embedded within parameter constraints and the order of optimisation steps, as well as the initial values for the fixed parameters. By explicitly presenting the trade-offs, the MOO framework empowers the user to make scientifically informed choices, rather than relying on the implicit biases inherent in traditional methods.

Finally, we emphasise that the \texttt{re:pair} framework is not designed to compete with nor replace the traditional analysis methods. In fact, the framework presented in this study offers a complementary alternative to methods that are typically employed for detailed analyses of individual binary systems. The power of the \texttt{re:pair} framework is in its objective exploration of the full parameter space and internally consistent way of treating different types of observational data. This makes the framework an ideal tool for studying stellar samples of intermediate sizes, especially in the context of global issues, such as the mass discrepancy problem, which require a good understanding of systematic uncertainties.

\section*{Data availability}\label{sec:data}

The framework developed in this work is available on GitHub and is open to the public (but it is still under active development); \href{https://github.com/PraiseTheCode/repair}{re:pair [ Revising Eclipsing binaries analysis : Photometry And spectroscopy Infused Recipe ]}. The repository includes both the main console application for multi-objective optimisation and the GUI for post-selection and post-processing.

Additional Appendices D–F containing remaining validation plots are available on Zenodo at \href{https://zenodo.org/records/15388159}{https://zenodo.org/records/15388159}, following the request of the Editor.

\begin{acknowledgements}

Based on observations made with the Mercator Telescope, operated on the island of La Palma by the Flemish Community, at the Spanish Observatorio del Roque de los Muchachos of the Instituto de Astrofísica de Canarias. Based on observations obtained with the {\sc hermes} spectrograph, which is supported by the Research Foundation - Flanders (FWO), Belgium, the Research Council of KU Leuven, Belgium, the Fonds National de la Recherche Scientifique (F.R.S.-FNRS), Belgium, the Royal Observatory of Belgium, the Observatoire de Genève, Switzerland and the Thüringer Landessternwarte Tautenburg, Germany. Funding provided by METH/24/012 Methusalem SOUL. 

The TESS data in this paper were obtained from the Mikulski Archive for Space Telescopes (MAST) at the Space Telescope Science Institute (STScI), which is operated by the Association of Universities for Research in Astronomy, Inc., under NASA contract NAS5-26555. Support to MAST for these data is provided by the NASA Office of Space Science via grant NAG5-7584 and by other grants and contracts. Funding for the TESS mission is provided by the NASA Explorer Program. 

This research has made use of the SIMBAD database, operated at CDS, Strasbourg, France; the SAO/NASA Astrophysics Data System; and the VizieR catalog access tool, CDS, Strasbourg, France.

\end{acknowledgements}

\bibliographystyle{aa}

\bibliography{aa53605-24.bib}

\begin{thebibliography}{43}
\expandafter\ifx\csname natexlab\endcsname\relax\def\natexlab#1{#1}\fi

\bibitem[{Budding {et~al.}(2009)Budding, Inlek, \& Demircan}]{budding_absolute_2009}
Budding, E., Inlek, G., \& Demircan, O. 2009, \mnras, 393, 501

\bibitem[{Catanzaro {et~al.}(2024)Catanzaro, Frasca, Alonso-Santiago, \& Colombo}]{catanzaro_tess_2024}
Catanzaro, G., Frasca, A., Alonso-Santiago, J., \& Colombo, C. 2024, \aap, 685, A133

\bibitem[{Claret \& Torres(2016)}]{claret_dependence_2016}
Claret, A. \& Torres, G. 2016, \aap, 592, A15

\bibitem[{Claret \& Torres(2017)}]{claret_dependence_2017}
Claret, A. \& Torres, G. 2017, \apj, 849, 18

\bibitem[{Claret \& Torres(2018)}]{claret_dependence_2018}
Claret, A. \& Torres, G. 2018, \apj, 859, 100

\bibitem[{Claret \& Torres(2019)}]{claret_dependence_2019}
Claret, A. \& Torres, G. 2019, \apj, 876, 134

\bibitem[{Deb {et~al.}(2002)Deb, Pratap, Agarwal, \& Meyarivan}]{deb_fast_2002}
Deb, K., Pratap, A., Agarwal, S., \& Meyarivan, T. 2002, Evolutionary Computation, IEEE Transactions on, 6, 182

\bibitem[{Fortin {et~al.}(2012)Fortin, De~Rainville, Gardner, Parizeau, \& Gagné}]{fortin_deap_2012}
Fortin, F.-A., De~Rainville, F.-M., Gardner, M., Parizeau, M., \& Gagné, C. 2012, Journal of Machine Learning Research, Machine Learning Open Source Software, 13, 2171

\bibitem[{{Guinan} {et~al.}(2000){Guinan}, {Ribas}, {Fitzpatrick}, {Gim{\'e}nez}, {Jordi}, {McCook}, \& {Popper}}]{Guinan2000}
{Guinan}, E.~F., {Ribas}, I., {Fitzpatrick}, E.~L., {et~al.} 2000, \apj, 544, 409

\bibitem[{{Hadrava}(1995)}]{Hadrava1995}
{Hadrava}, P. 1995, \aaps, 114, 393

\bibitem[{Herrero(2007)}]{herrero_spectroscopic_2007}
Herrero, A. 2007, Highlights of Astronomy, 14, 201

\bibitem[{Herrero {et~al.}(2000)Herrero, Puls, \& Villamariz}]{herrero_fundamental_2000}
Herrero, A., Puls, J., \& Villamariz, M.~R. 2000, \aap, 354, 193

\bibitem[{Holmgren {et~al.}(1991)Holmgren, Hill, \& Fisher}]{holmgren_absolute_1991}
Holmgren, D.~E., Hill, G., \& Fisher, W. 1991, \aap, 248, 129

\bibitem[{Ilijic {et~al.}(2004)Ilijic, Hensberge, Pavlovski, \& Freyhammer}]{ilijic_obtaining_2004}
Ilijic, S., Hensberge, H., Pavlovski, K., \& Freyhammer, L.~M. 2004, ASP Conference Series, 318, 111

\bibitem[{Ilijić(2017)}]{ilijic_fd3_2017}
Ilijić, S. 2017, Astrophysics Source Code Library, ascl:1705.012

\bibitem[{Johnston {et~al.}(2019)Johnston, Pavlovski, \& Tkachenko}]{johnston_modelling_2019}
Johnston, C., Pavlovski, K., \& Tkachenko, A. 2019, \aap, 628, A25

\bibitem[{Khan \& Shulyak(2006{\natexlab{a}})}]{khan_stellar_2006}
Khan, S.~A. \& Shulyak, D.~V. 2006{\natexlab{a}}, \aap, 448, 1153

\bibitem[{Khan \& Shulyak(2006{\natexlab{b}})}]{khan_stellar_2006-1}
Khan, S.~A. \& Shulyak, D.~V. 2006{\natexlab{b}}, \aap, 454, 933

\bibitem[{Kochukhov {et~al.}(2005)Kochukhov, Khan, \& Shulyak}]{kochukhov_stellar_2005}
Kochukhov, O., Khan, S., \& Shulyak, D. 2005, \aap, 433, 671

\bibitem[{Kurucz(1979)}]{kurucz_model_1979}
Kurucz, R.~L. 1979, \apjs, 40, 1

\bibitem[{{Lightkurve Collaboration} {et~al.}(2018){Lightkurve Collaboration}, {Cardoso}, {Hedges}, {Gully-Santiago}, {Saunders}, {Cody}, {Barclay}, {Hall}, {Sagear}, {Turtelboom}, {Zhang}, {Tzanidakis}, {Mighell}, {Coughlin}, {Bell}, {Berta-Thompson}, {Williams}, {Dotson}, \& {Barentsen}}]{lightkurve}
{Lightkurve Collaboration}, {Cardoso}, J. V. d.~M., {Hedges}, C., {et~al.} 2018, {Lightkurve: Kepler and TESS time series analysis in Python}, Astrophysics Source Code Library, record ascl:1812.013

\bibitem[{Markova {et~al.}(2018)Markova, Puls, \& Langer}]{markova_spectroscopic_2018}
Markova, N., Puls, J., \& Langer, N. 2018, \aap, 613, A12

\bibitem[{Maxted(2016)}]{maxted_ellc_2016}
Maxted, P. F.~L. 2016, \aap, 591, A111

\bibitem[{Pavlovski {et~al.}(2018)Pavlovski, Southworth, \& Tamajo}]{pavlovski_physical_2018}
Pavlovski, K., Southworth, J., \& Tamajo, E. 2018, \mnras, 481, 3129

\bibitem[{{Pavlovski} {et~al.}(2009){Pavlovski}, {Tamajo}, {Koubsk{\'y}}, {Southworth}, {Yang}, \& {Kolbas}}]{Pavlovski2009}
{Pavlovski}, K., {Tamajo}, E., {Koubsk{\'y}}, P., {et~al.} 2009, \mnras, 400, 791

\bibitem[{Prša {et~al.}(2016)Prša, Conroy, Horvat, Pablo, Kochoska, Bloemen, Giammarco, Hambleton, \& Degroote}]{prsa_physics_2016}
Prša, A., Conroy, K.~E., Horvat, M., {et~al.} 2016, \apjs, 227, 29

\bibitem[{Prša \& Zwitter(2005)}]{prsa_computational_2005}
Prša, A. \& Zwitter, T. 2005, \apj, 628, 426

\bibitem[{{Raskin} {et~al.}(2011){Raskin}, {van Winckel}, {Hensberge}, {Jorissen}, {Lehmann}, {Waelkens}, {Avila}, {de Cuyper}, {Degroote}, {Dubosson}, {Dumortier}, {Fr{\'e}mat}, {Laux}, {Michaud}, {Morren}, {Perez Padilla}, {Pessemier}, {Prins}, {Smolders}, {van Eck}, \& {Winkler}}]{Raskin2011}
{Raskin}, G., {van Winckel}, H., {Hensberge}, H., {et~al.} 2011, \aap, 526, A69

\bibitem[{{Simon} \& {Sturm}(1994)}]{Simon1994}
{Simon}, K.~P. \& {Sturm}, E. 1994, \aap, 281, 286

\bibitem[{Southworth {et~al.}(2020)Southworth, Bowman, Tkachenko, \& Pavlovski}]{southworth_discovery_2020}
Southworth, J., Bowman, D.~M., Tkachenko, A., \& Pavlovski, K. 2020, \mnras, 497, L19

\bibitem[{Southworth {et~al.}(2004)Southworth, Maxted, \& Smalley}]{southworth_eclipsing_2004}
Southworth, J., Maxted, P. F.~L., \& Smalley, B. 2004, \mnras, 351, 1277

\bibitem[{Southworth {et~al.}(2005)Southworth, Smalley, Maxted, Claret, \& Etzel}]{southworth_absolute_2005}
Southworth, J., Smalley, B., Maxted, P. F.~L., Claret, A., \& Etzel, P.~B. 2005, \mnras, 363, 529

\bibitem[{Srinivas \& Deb(1994)}]{srinivas_muiltiobjective_1994}
Srinivas, N. \& Deb, K. 1994, Evolutionary Computation, 2, 221

\bibitem[{Takeda {et~al.}(2019)Takeda, Han, Kang, Lee, \& Kim}]{takeda_compositional_2019}
Takeda, Y., Han, I., Kang, D.-I., Lee, B.-C., \& Kim, K.-M. 2019, \mnras, 485, 1067

\bibitem[{Tkachenko(2015)}]{tkachenko_grid_2015}
Tkachenko, A. 2015, \aap, 581, A129

\bibitem[{{Tkachenko} {et~al.}(2014){Tkachenko}, {Degroote}, {Aerts}, {Pavlovski}, {Southworth}, {P{\'a}pics}, {Moravveji}, {Kolbas}, {Tsymbal}, {Debosscher}, \& {Cl{\'e}mer}}]{Tkachenko2014}
{Tkachenko}, A., {Degroote}, P., {Aerts}, C., {et~al.} 2014, \mnras, 438, 3093

\bibitem[{Tkachenko {et~al.}(2020)Tkachenko, Pavlovski, Johnston, Pedersen, Michielsen, Bowman, Southworth, Tsymbal, \& Aerts}]{tkachenko_mass_2020}
Tkachenko, A., Pavlovski, K., Johnston, C., {et~al.} 2020, \aap, 637, A60

\bibitem[{Tkachenko {et~al.}(2024)Tkachenko, Pavlovski, Serebriakova, Bowman, IJspeert, Gebruers, \& Southworth}]{tkachenko_observational_2024}
Tkachenko, A., Pavlovski, K., Serebriakova, N., {et~al.} 2024, \aap, 683, A252

\bibitem[{{Torres} {et~al.}(2010){Torres}, {Andersen}, \& {Gim{\'e}nez}}]{Torres2010}
{Torres}, G., {Andersen}, J., \& {Gim{\'e}nez}, A. 2010, \aapr, 18, 67

\bibitem[{Tsymbal(1996)}]{tsymbal_starsp_1996}
Tsymbal, V. 1996, ASP Conference Series, 108, 198

\bibitem[{Vaz {et~al.}(2007)Vaz, Andersen, \& Claret}]{vaz_absolute_2007}
Vaz, L. P.~R., Andersen, J., \& Claret, A. 2007, \aap, 469, 285

\bibitem[{{Wilson} \& {Devinney}(1971)}]{Wilson-Devinney1971}
{Wilson}, R.~E. \& {Devinney}, E.~J. 1971, \apj, 166, 605

\bibitem[{Wilson \& Van~Hamme(2009)}]{wilson_distances_2009}
Wilson, R.~E. \& Van~Hamme, W. 2009, \apj, 699, 118

\end{thebibliography}

\begin{appendix}
\onecolumn
\newpage

\section{Uncertainties from confidence ellipses}\label{app:conf_ell}

\begin{figure*}[h]
   \centering
\includegraphics[clip,width=180mm,trim={0cm 0cm 0cm 0cm}]{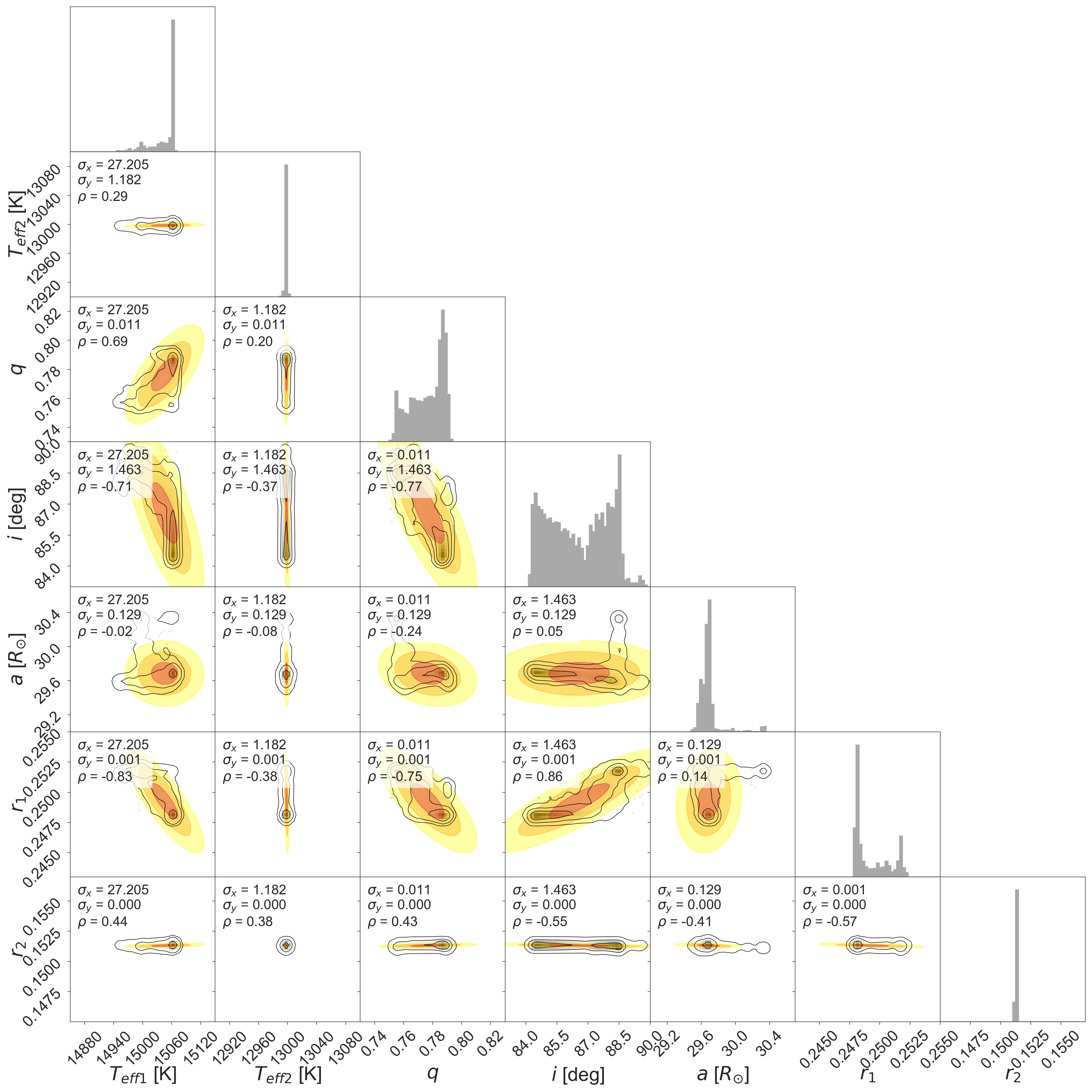}  
      \caption{ Confidence ellipses on top of density levels for the last generations of a test dataset. Each subplot below the diagonal contains confidence ellipses corresponding to $1\sigma$, $2\sigma$, and $3\sigma$ confidence levels (red, orange, and yellow, respectively), derived from the covariance matrix of the parameters. Annotations in each subplot indicate the standard deviations along the axes ($\sigma_x$ and $\sigma_y$) and the Pearson correlation coefficient ($\rho$). 
 }
    \label{Fig:B5_conf_ell}
\end{figure*}
\clearpage

\section{Observed spectroscopic time series of U Oph and V453 Cyg. }

\begin{figure}[!htp]
   \centering
\includegraphics[clip,width=500pt,trim={0.3cm 1cm 0.3cm 0.3cm}]{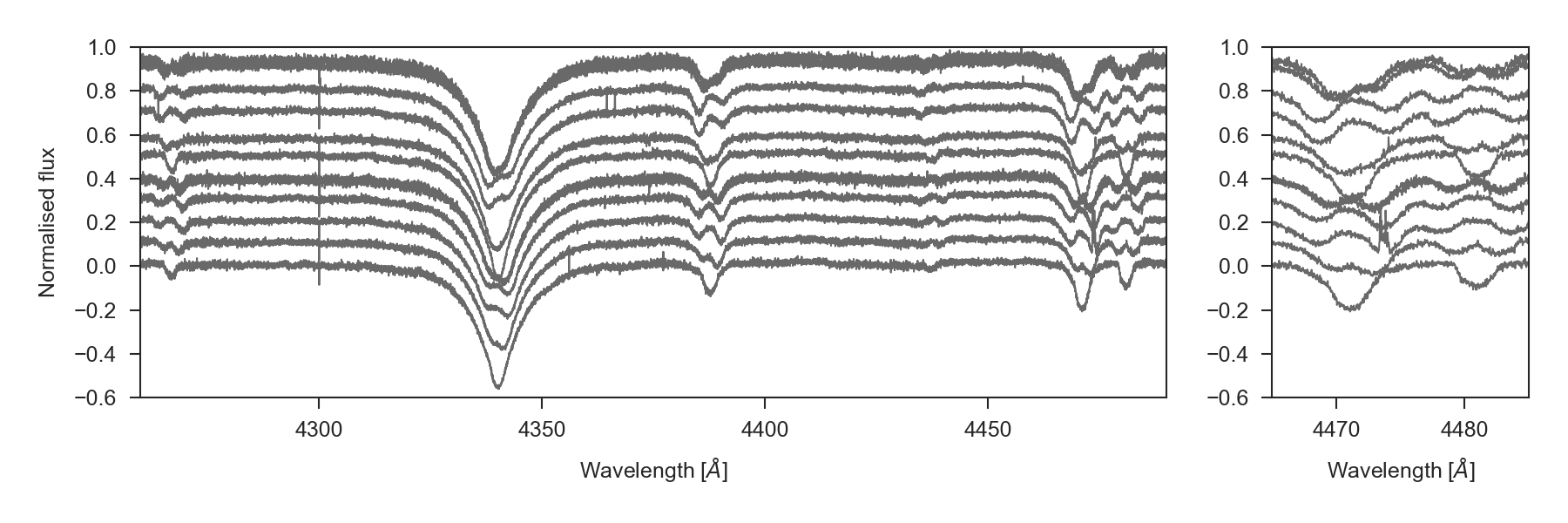}  
\includegraphics[clip,width=500pt,trim={0.3cm 0cm 0.3cm 0.3cm}]{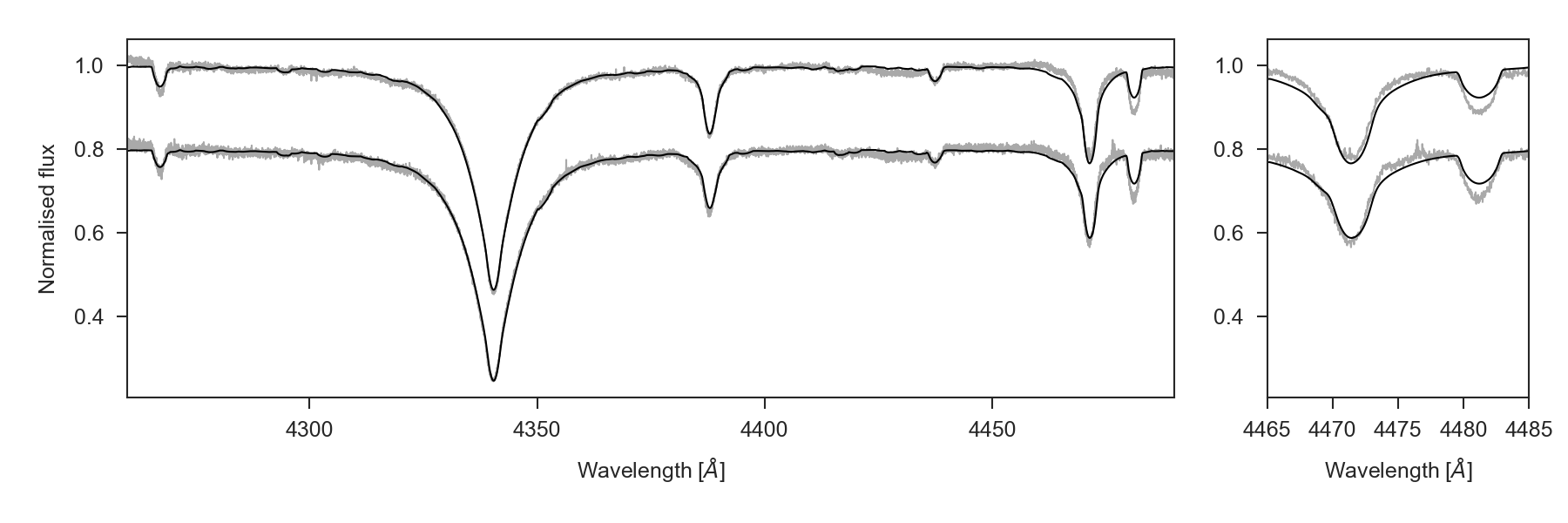} 
      \caption{Same as in Figure~\ref{Fig:WWAur_obsSpec}, but for the U Oph system. Solar composition was adopted.} 
    \label{Fig:UOph_obsSpec}
\end{figure}

\begin{figure}[!htp]
   \centering
\includegraphics[clip,width=500pt,trim={0.25cm 0.9cm 0.1cm 0.2cm}]{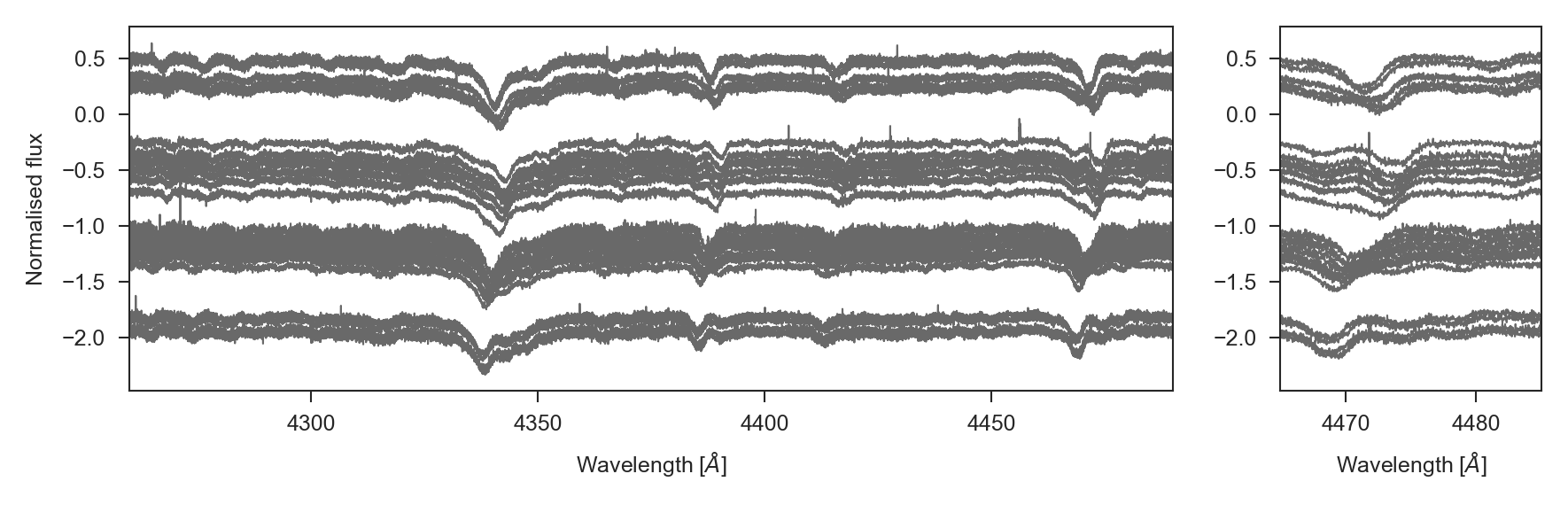}  
\includegraphics[clip,width=500pt,trim={0.1cm 0cm 0.25cm 0.2cm}]{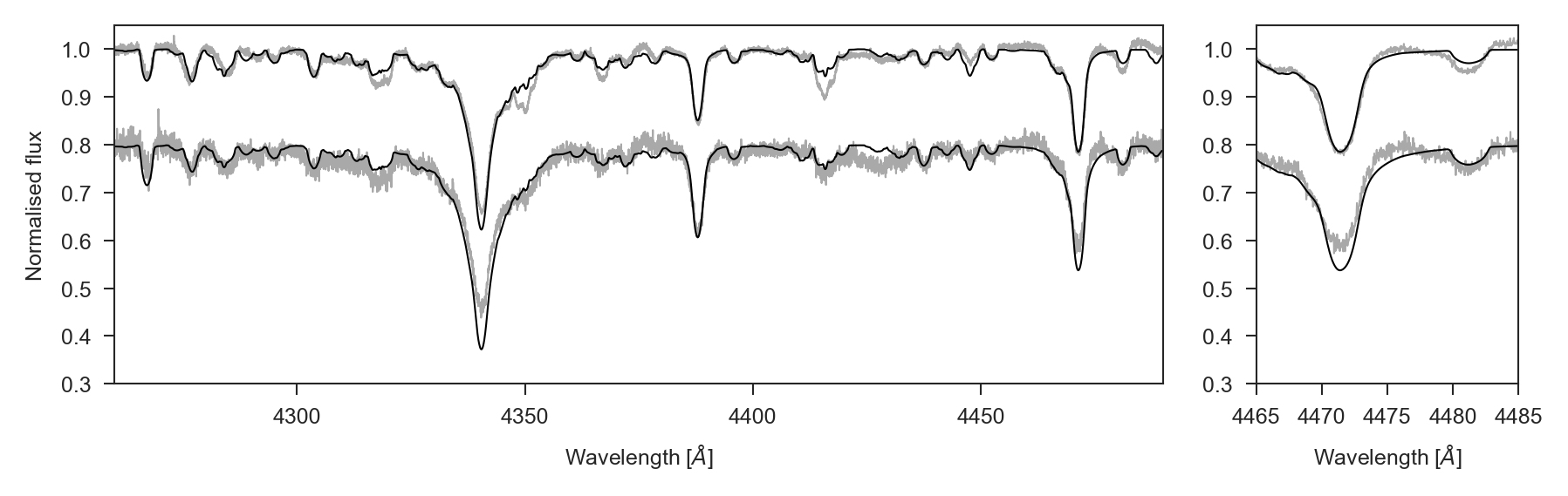}  
      \caption{Same as in Figure~\ref{Fig:WWAur_obsSpec}, but for the V453 Cyg system.} 
    \label{Fig:V453Cyg_obsSpec}
\end{figure}

\clearpage

\section{ An application for analysis of optimisation results. }

\begin{figure}[!htp]
   \centering 
\includegraphics[clip,width=169mm,trim={1.5cm 0.95cm 1.5cm 0cm}]{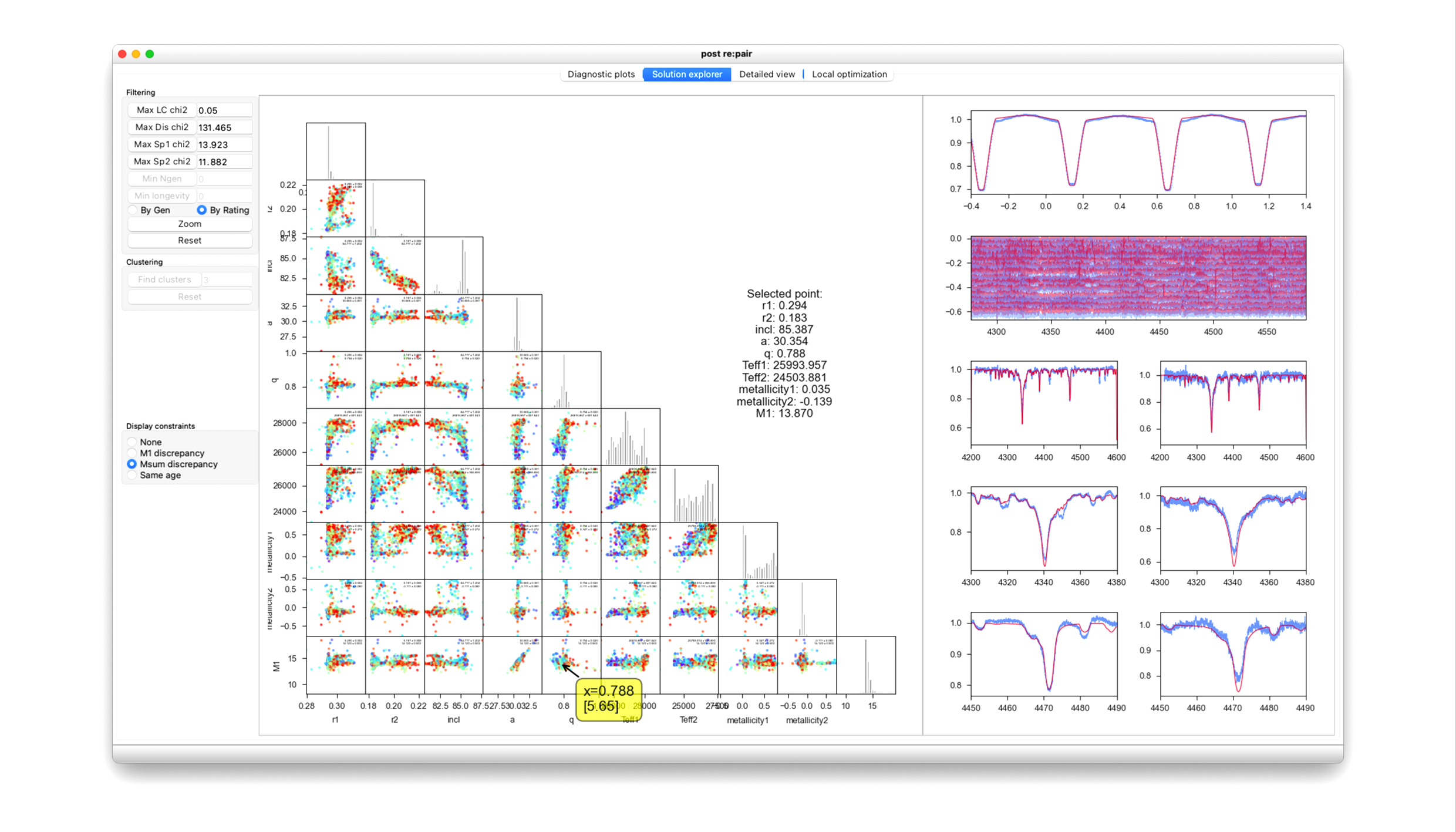}  

\includegraphics[clip,width=169mm,trim={1.5cm 0.95cm 1.5cm 0cm}]{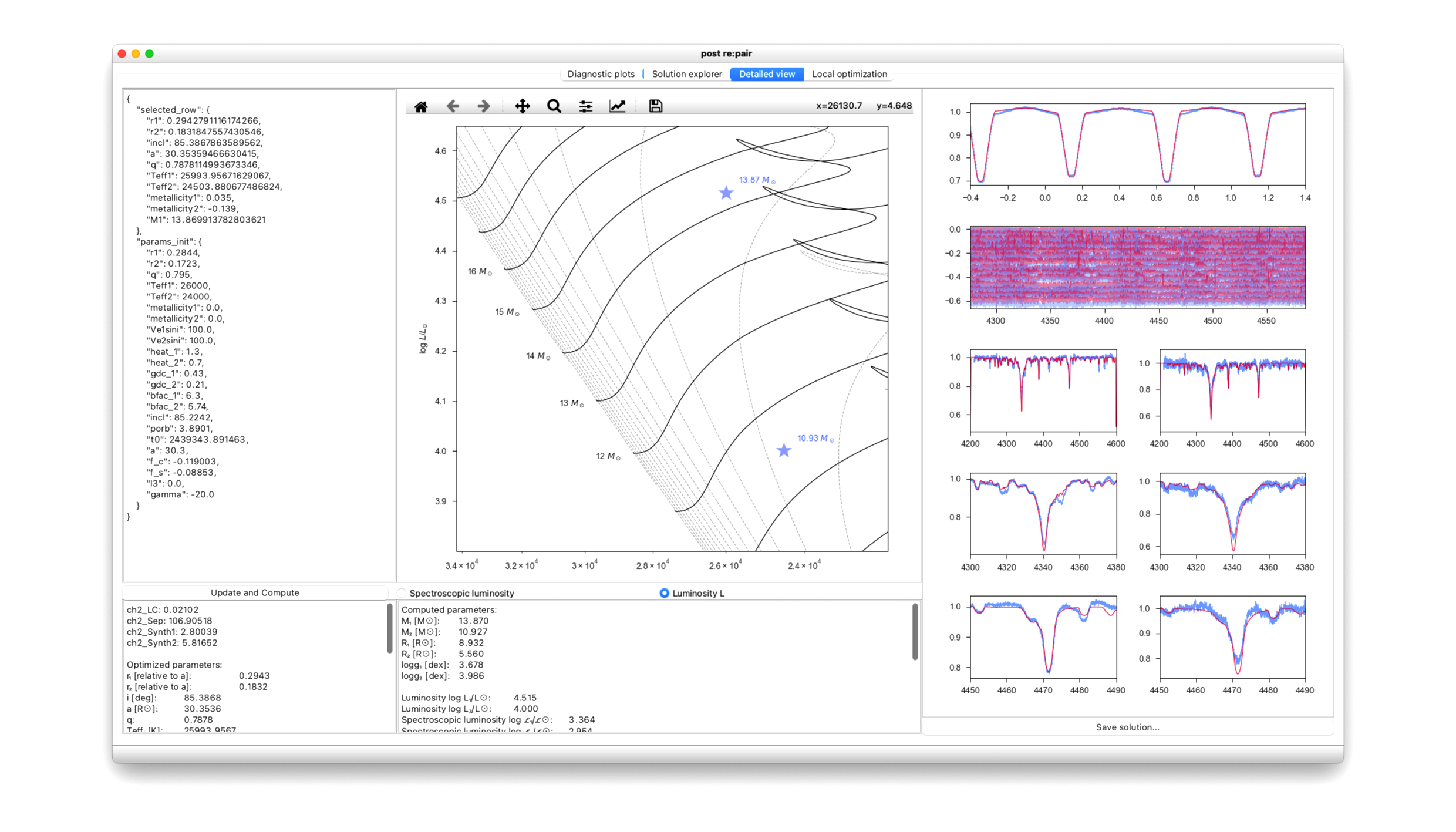}  

       \caption{Screenshots of the post-processing GUI. Top panel: a model-informed solution selection during post-process after multi-objective optimisation. Optional colour-coding according to the mass discrepancy and age difference allows for the determination of the region preferred by the model in the parameter space. An interactive selection of any point of interest allows us to compute the corresponding data fit on-the-fly: the light curve, spectral disentangling, and spectral fitting. Bottom panel: Detailed view of any selected point, that also allows for manual variation of the parameters or an automated local weighted optimisation (in a separate tab).   }
    \label{Fig:screen1-2}
\end{figure}

\end{appendix}

\end{document}